\documentstyle[12pt]{article}
\newcommand{\fslash}[1]{\mbox{$\!\not\!#1$}}
\newcommand{\Lag}{{\cal L}}

\newcommand{\be}{\begin{equation}}
\newcommand{\ee}{\end{equation}}

\newcommand{\bold}[1]{\mbox{\boldmath ${#1}$}}

\newcommand{\kint}[1]{\int \! \!{d^4 #1 \over {(2\pi)^4}}}

\newcommand{\refb}[1]{(\ref{#1})}

\begin{document}

\baselineskip 4 ex

\title{NJL model on the light cone and pion structure function
\thanks{Correspondence to: W. Bentz, E-mail: bentz@tkynt2.phys.s.u-tokyo.
ac.jp}}
       
\author{W. Bentz, T. Hama, T. Matsuki, K. Yazaki  \\
        Department of Physics, \\
        Faculty of Science \\
        University of Tokyo \\
        Hongo 7-3-1, Bunkyo-ku, Tokyo 113, Japan}

\date{}
\maketitle
\begin{abstract}
The NJL model is formulated on the light cone. Using the 1/N expansion
to solve the fermionic constraint,
an effective 4-fermi lagrangian is derived. Pionic properties are investigated
using this lagrangian, and the formal equivalence to the equal-time
formulation is demonstrated. Two regularization schemes in terms of
light-cone variables are discussed: An extension of the 'invariant mass
cut-off scheme', and a transverse momentum cut-off scheme. It is shown
that the first one is equivalent to the 
covariant 3-momentum cut-off (dispersion cut-off) scheme in the equal-time 
formulation. As an application the structure function of the pion is
studied in both regularization schemes. \\ \\ \\
{\footnotesize PACS numbers: }
\end{abstract}

\newpage

\section{Introduction}
\setcounter{equation}{0}

As an effective theory of QCD in the low energy region,
the Nambu - Jona-Lasinio (NJL) model \cite{NJL} is a powerful 
tool to investigate
the properties of hadrons \cite{HADR}. It exhibits the spontaneous breaking of
chiral symmetry and the appearance of Goldstone bosons in a simple and
clear way. Due to the simplicity of the interaction,
the relativistic two- and three-body equations can be solved in the
ladder approximation, which allows a consistent unified description
of mesons and baryons \cite{FAD1,FAD2}. Recently, extensions have been made
to include also mesonic degrees of freedom into the description of
three-quark bound states \cite{ISH}, and methods to incorporate the effects of
confinement phenomenologically have been proposed \cite{CONF}. Concerning the
ground state properties of baryons, the NJL model has been quite successful
\cite{GROUND}.
 
Recently, the model has been used also to describe the structure functions
of the pion \cite{PIONS} and the nucleon \cite{NUCLS} measured in deep 
inelastic scattering.
It is well known that in the Bjorken limit the structure functions
are expressed as Fourier transforms of correlation functions
along the light-cone (LC) \cite{JAFFE}. 
Therefore, for the evaluation of the structure functions in the
Bjorken limit the use of LC variables is most natural \cite{BURK}. 
The interpretation of
the structure functions as LC momentum distributions of the
constituents (partons) in the ground state of the hadron, however,
is strictly speaking possible only in a framework where 
the LC 
correlation functions can be considered as equal 'time' correlations.
In the LC quantization procedure \cite{BURK,YAN,YAZ} this is indeed the case, 
since in this approach the fields
evolve in the 'time' variable $x^+=(x^0+x^3)/\sqrt{2}\,$ 
instead of the usual time variable $x^0$.
Therefore, from the physical standpoint one expects that the most natural 
way to describe the structure
functions in the Bjorken limit consists of not only using 
LC {\em variables}, but also LC
{\em quantization}. The LC quantization, which was proposed by Dirac 
\cite{DIR} a long time ago,
has also many other interesting features, like the 'vacuum triviality' 
\cite{LEU} and
associated with this the fact that Fock space expansions of physical
states have an exact meaning, without referring to a particular approximation
scheme \cite{FOCK}. 

In spite of these physically appealing features, LC treatments also have
$-$ mostly technical $-$ disadvantages. For example, they conceal the underlying
rotational invariance and complicate the treatment of angular momentum,
and bring in new problems of how to regularize divergent
integrals \cite{ITA}. Moreover, the presence of zero modes
destroys the 'vacuum triviality' in the naive sense, and the proper
treatment of these zero modes is actually very complicated and technically
involved \cite{ZERO}. Therefore, in actual calculations of structure functions
in the NJL model, it might eventually be more convenient to use
the usual equal time (ET) quantization and merely perform a  
transformation
to LC variables at some convenient stage. Nevertheless, it is an
important task to show that such a procedure is indeed equivalent to the
LC quantization method with all its physical merits. To investigate this
equivalence for the case of the NJL model is the purpose of this work. 

The demonstration of the equivalence of ET
and LC quantization for $\sigma$-model type theories was the subject
of previous investigations \cite{YAN}. For 4-fermi theories like the NJL
model, it is well
known that the LC quantization poses additional problems due to the
nonlinearity of the fermionic constraint \cite{ITA,REG}. 
To circumvent this problem, 
in some previous 
works \cite{REG} this fermionic constraint equation was replaced by the 
corresponding 
equation for free fields. In this work we will show a straight forward 
method \cite{HAMA} based on the $1/N$ expansion up to the next-to-leading 
order to solve the fermionic constraint. (For methods using boson
mapping techniques see ref. \cite{ITA}.) 
We will show that the resulting effective lagrangian has the same form
as obtained by using the free field equations to solve the fermionic
constraint, but the 4-fermi coupling constant is replaced by an
effective one. 
For physical quantities, the changes due to this modified coupling
constant cancel against those due to modified propagators, such that
the theory is formally equivalent to the ET theory. We will show that
there exist cut-off schemes where this equivalence holds even
for the regularized theories. Namely, a suitably extended version 
of the "invariant mass cut-off" introduced
in ref. \cite{LB} will be shown to be equivalent to the 
ET treatment based on
the covariant 3-momentum cut-off (dispersion cut-off) \cite{HK}. 
We will also discuss a recently proposed alternative regularization scheme 
\cite{TRAN} which has no
counterpart in the conventional ET treatment.
As an application, we will discuss the pion structure function,
and compare our results with the empirically derived valence quark 
distributions in the pion \cite{EMP}.

The rest of this paper is organized as follows: In sect. 2 we review
some well known facts of the NJL model in the ET formulation based
on the covariant 3-momentum cut-off scheme, and discuss the pion structure
function in the ET formulation. In sect. 3 we derive the effective
LC lagrangian in the framework of the 1/N expansion, discuss the
new Feynman rules based on the LC quantization, and show the formal
equivalence to the ET theory. Two regularization schemes suitable for
LC theories will be discussed. In sect. 4 
we will show our numerical results for the valence quark LC momentum 
distribution in the pion, and in sect. 5 we summarize our work.

\section{NJL model in the ET quantization and the pion structure function}
\setcounter{equation}{0}

\subsection{The model}

The lagrangian of the flavor SU(2) NJL model in its original form 
is given by \cite{NJL}
\be
\Lag = \Lag_0 + \Lag_I,
\label{lag}
\ee
where $\Lag_0 = \overline{\psi}(i\fslash{\partial} - m)\psi$ is
the free quark lagrangian with $m$ the current $u$, $d$ quark mass,
and $\Lag_I$ is given by
\be
{\cal L}_I = G
\biggl( \Bigl(\overline{\psi}\psi \Bigr)^2
	- \Bigl(\overline{\psi}(\gamma_5\bold{\tau})\psi\Bigr)^2
\biggr). \label{lagi}
\ee
The gap equation is given as usual by
\be
M = m - 2 G \langle {\overline \psi} \psi \rangle = 
m + 12iG\kint{q}tr_D\Bigl(S(q)\gamma^0). \label{gap}
\ee
$M$ is the constituent quark mass, 
${\displaystyle S(q) = {1 \over \fslash{q} - M+i\epsilon}\gamma^0}$ is 
the Feynman propagator
for the constituent quark \footnote{To correspond to the conventions used
in most papers on LC field theories (see for example ref. \cite{BURK}), 
we define the Feynman propagator
via the T-product of $\psi$ and $\psi^+$, and normalize spinors according
to $u^{\dagger}({\bold p},s') u({\bold p},s)=2 E_p \delta_{s' s}$. 
We also restrict
ourselves to the lowest order in the $1/N$ expansion (direct terms only), 
and therefore the effective coupling
constant in the scalar and pseudoscalar channels is given by $G$ (instead of
$\frac{13}{12} G$). Throughout this paper,  
$<{\overline \psi} \Gamma \psi>$ denotes the leading order ($N$) of the
vacuum expectation value of ${\overline \psi} \Gamma \psi$.}, 
and $\langle {\overline \psi} \psi \rangle$
denotes the quark condensate in the leading order ($N$) of the 
$1/N$ expansion.
The solution of the BS equation for the $q {\overline q}$ t-matrix 
in the pionic channel is
\be
t(k)=(\gamma^0 \gamma_5\tau_i) \tau(k)(\gamma ^0 \gamma_5\tau_i).
\label{tpi}
\ee
Here the reduced t-matrix is given by
\be
\tau(k)={-2iG \over 1 + 2 G \Pi(k)},
\label{taupi}
\ee
where
\be
\Pi(k)=6 i \kint{q} tr_D \biggl[
	\gamma^0 \gamma_5 S(q) \gamma^0 \gamma_5  S(q-k)\biggr]
\label{bubb}
\ee
is the familiar 'bubble graph'.
The pion mass $m_{\pi}$ is obtained as the pole of \refb{taupi}.

The Bethe-Salpeter (BS) wave function $\Phi_k(q)$ in the covariant 
normalization is defined 
via the residue of the connected $q {\overline q}$ Green function 
at the pion pole $k_0=\omega_k \equiv \sqrt{{\bold k}^2+m_{\pi}^2}$, i.e;
\be
S(q')S(q'-k) t(k) S(q-k) S(q) \longrightarrow
\frac{i}{2\omega_k} \frac{\Phi^i_k(q') \Phi^{i+}_k(q)}{k^0-\omega_k+i \epsilon}
\,\,\,\,\,\,\,\,\,{\rm as}\,\,k_0\rightarrow \omega_k,
\label{lim}
\ee
which leads to
\be
\Phi_k^i(q)=g \biggl(S(q) \gamma^0 \gamma_5 \tau_i S(q-k)\biggr);
\,\,\,\,\,\,\,\,\,\,\,\,\,
\frac{1}{g^2}=-\left(\frac{\partial \Pi}{\partial k^2}\right)_{k^2=m_{\pi}^2}.
\label{wave}
\ee
Here $g$ is the pion-quark coupling constant. The assignment of the external
momenta in eq. \refb{lim} is shown in fig. \ref{mom}. 

The pion decay constant is defined as usual via the matrix element of
the axial vector current between the pion state $|ki>$ and the vacuum:
\be
<0|\overline{\psi}(0)\frac{\tau_j}{2}\gamma^{\mu}\gamma_5 \psi(0)|ki>=
i\delta_{ij} k^{\mu}f_{\pi}
\label{dfpi}
\ee
This amounts to calculating the familiar one-quark loop contribution to the
pion decay, using the BS wave function
eq. \refb{wave} at the ${\overline q} q \pi$ vertex. The results is 
\begin{eqnarray}
i k^{\mu} f_{\pi} &=& 3 g \kint{q} tr_D\left(\gamma^0 \gamma^{\mu} \gamma_5
S(q-k) \gamma^0 \gamma_5 S(q)\right) \label {loop}\\ 
&=& 12 g M k^{\mu} \kint{q} \frac{1}{\left(q^2-M^2+i\epsilon\right)
\left( \left(
q-k\right)^2-M^2+i\epsilon\right)} \,\,\,\,\,\,\,\,\,(k^2=m_{\pi}^2).
\nonumber \\
\label{fpi}
\end{eqnarray}

\subsection{Quark LC momentum distribution}

In the ET quantization method, the 
valence quark LC momentum distribution $q_v(x)$ in the pion as a function 
of the Bjorken variable $x$ \cite{JAFFE} 
\be
q_v(x) = \frac{1}{4\pi} \int dz^- e^{ixk_-z^-} \langle ki|{\overline \psi}(0)
\gamma^+ \psi(z^-)|ki \rangle \label{qv1} 
\label{qv2}
\ee  
can be obtained from the
diagram fig. \ref{dia}, where the external operator is $\gamma^+$ and the loop
integral is performed with respect to $q_+$ and ${\bf q}_{\perp}$ with
$q_-=x k_-$ fixed \footnote{Our conventions for LC variables
are summarized in Appendix A.}. Using $S(q)\gamma^0 \gamma^+ S(q) =
- \partial S(q)/\partial q_+$  
and performing a partial integration in $q_+$, we get
\begin{eqnarray}
q_v(x)&=&3ig^2 \int \frac{dq_+ d^2q_{\perp}}{(2\pi)^4} 
tr_D\biggl(S(q) \gamma^0 \gamma^+ S(q) \gamma^0 \gamma_5 S(q-k) \gamma^0
\gamma_5\biggr) \label{loop1} \\
&=& - g^2 \frac{\partial \Pi(k,x)}{\partial k^2} \,\,\,\,\,\,\,\,
(k^2=m_{\pi}^2),
\label{qv}
\end{eqnarray}
where
\be
\Pi(k,x)=6 i k_-  \int \frac{dq_+ d^2q_{\perp}}{(2\pi)^4} 
tr_D\biggl(\gamma^0 \gamma_5 S(q) \gamma^0 \gamma_5 S(q-k)\biggr)
\label{pix}
\ee
with ${\displaystyle x=\frac{q_-}{k_-}}$. The quantity \refb{pix} 
is defined such that 
\be
\int_{-\infty} ^{\infty} dx \Pi(k,x)=\Pi(k).
\label{inte}
\ee
To evaluate \refb{pix}, we need the Feynman propagator in terms of LC
variables: 
\be
S(p)=\frac{\gamma^+p_+ + \gamma^- p_- - {\bold \gamma}_{\perp}\cdot
{\bold p}_{\perp}+M}{p^2-M^2 +i \epsilon} \gamma^0 \,\,,
\label{fey} 
\ee
where 
\be
\frac{1}{p^2-M^2+i\epsilon} = \frac{1}{2 p_-} \left(\frac{\Theta(p_-)}
{p_+ - e_p + i\epsilon} + \frac{\Theta(-p_-)}{p_+ - e_p - i \epsilon}
\right)
\label{den}
\ee
with
\be
e_p = \frac{{\bold p}_{\perp}^2+M^2}{2p_-}.
\label{eps}
\ee
The Feynman propagator \refb{fey} contains a contact term
in the LC 'time' variable $x^+$ (conjugate momentum $p_+$), and it is
convenient to separate this term:
\begin{equation}
S(p) = \frac{\overline {\fslash{p}}+M}{p^2-M^2+i\epsilon} \gamma^0
+ \frac{\gamma^+ \gamma^0}{2 p_-} \equiv  
{\overline S}(p) + \frac{\gamma^+ \gamma^0}{2 p_-}, 
\label{sbar}
\end{equation}
where ${\overline p}^{\mu}$ is the on-shell four momentum with LC
components $(p_+= e_p, p_-, {\bold p}_{\perp})$.
Inserting this into eq. \refb{pix} and \refb{inte}, one obtains
\be 
\Pi(k,x) = {\overline \Pi}(k,x) + R(x), \,\,\,\,\,\,\,\,\,\,\,\,\,\,\,\,\,
\Pi(k)={\overline \Pi}(k) + R,
\label{piece}
\ee
where 
${\overline \Pi}(k,x)$ $\,\,\left({\overline \Pi}(k)\right)\,\,$ 
is obtained by 
replacing $S\rightarrow {\overline S}$
in eq. \refb{pix} (eq. \refb{bubb}). For ${\bold k}_{\perp}=0$ we obtain 
\begin{eqnarray}
{\overline \Pi}(k,x)&=&- \frac{6 \Theta(x) \Theta(1-x)}{x(1-x)} 
\int \frac{d^2 q_{\perp}}
{\left(2\pi\right)^3} \frac{{\bold q}_{\perp}^2+M^2}{{\bold q}_{\perp}^2+M^2
-k^2x(1-x)-i\epsilon}  \nonumber \\ 
\label{pibarx} \\
{\overline \Pi}(k)&=&\int_0^1 dx \,{\overline \Pi}(k,x),
\label{pibar}
\end{eqnarray}
while $R$ is due to the contact term and given by
\be
R = \int_{-\infty}^{\infty} dx\,R(x) =
-3 i {\cal P} \kint{q} 
tr_D \left( \frac{{\overline S}(q)\gamma^0 \gamma^+}
{q_--k_-} + \frac{{\overline S}(q-k)\gamma^0 \gamma^+}{q_-}\right), 
\label{rx1} 
\ee
where ${\cal P}$ denotes the principal value with respect to the
$q_-$ integration.
Due to eq. \refb{sbar}, ${\overline S}$ can be replaced by $S$
in this expression. Using \refb{den} and performing the $q_+$ integral
we obtain
\footnote{We define 
\begin{eqnarray}
\int_{-\infty}^{\infty} \frac{dx}{x-a\pm i\epsilon}
&\equiv& \lim _{R\rightarrow \infty} \int_{-R}^{R} \frac{dx}{x-a\pm i\epsilon}
= \mp i\pi \nonumber
\end{eqnarray}
where the second equality follows from usual contour integration, taking
into account the contribution from the semicircle. 
}
\be
R= -3 {\cal P} \int_{-\infty}^{\infty} dx \left(\frac{\Theta(x)}
{x-1}-\frac{\Theta(-x)}{x-1} 
+ \frac{\Theta(x+1)}{x} - \frac{\Theta(x-1)}{x}\right)
\int \frac{d^2 q_{\perp}}{\left(2\pi\right)^3} . \\ 
\label{rx2}
\ee
It should be noted that in obtaining the expressions \refb{pibarx} and
\refb{rx2}, no shifts of integration variables have been performed.

The distribution function \refb{qv} then becomes
\be
q_v(x)= -g^2 \frac{\partial {\overline \Pi}(k,x)}{\partial k^2} =  
6 g^2 \int \frac{d^2 q_{\perp}}{(2\pi)^3} 
\frac{{\bold q}_{\perp}^2+M^2}{\left({\bold q}_{\perp}^2+M^2
-m_{\pi}^2 x (1-x)\right)^2},
\label{result}
\ee
and since $1/g^2=-\partial {\overline \Pi}(k)/\partial k^2$ $(k^2=m_{\pi}^2$) 
(see eq. \refb{wave} and \refb{piece}), the sum rules
\begin{equation}
\int_0^1 dx\, q_v(x) =1 ; \,\,\,\,\,\,\,\,\,\,
\int_0^1 dx \,x\, q_v(x) = \frac{1}{2}
\label{sumr}
\end{equation} 
follow immediately. The physical content of these relations is that we have
one valence quark in the pion which carries half of the LC momentum
$k_-$.   

Finally, from the following identities 
\be
-i \int \frac{d^4 q}{\left(2\pi\right)^4} \frac{1}{\left(q^2-\mu^2+i\epsilon
\right)^2} = 
\frac{1}{2} \int dq_- \int \frac{d^2 q_{\perp}}{\left(2 \pi \right)^3}
\frac{\delta(q_-)}{{\bold q}_{\perp}^2+\mu^2}=
\frac{1}{2} \int \frac{d^2 q_{\perp}}{\left(2 \pi \right)^3}
\frac{1}{{\bold q}_{\perp}^2+\mu^2}
\label{ide}
\ee
we note that:
(i) One could rewrite the expression \refb{result}
for $q_v(x)$ in terms
of a 4-dimensional integral, which agrees with the form given in 
previous works \cite{PIONS}. Such a procedure could be used in order to
avoid the regularization in terms of LC coordinates, but we do not
follow it in this paper. (ii) There is no contribution from the vertex
correction shown in fig.\ref{vertex}: This graph, which arises from 
the structure
of the constituent quark itself, gives a contribution $\propto \delta(x)$
when inserted into the graph of fig.\ref{dia}. Due to the regularization 
procedure discussed below, this gives no contribution.

\subsection{Regularization}

In the present ET formulation, we will regularize
the gap equation \refb{gap}, the bubble graph \refb{bubb} and the expression
for the pion
decay constant \refb{fpi} using the covariant 3-momentum cut-off ($\Lambda_3$)
as explained in ref. \cite{HK}. (This scheme will be denoted by '3M'.)
Basically, this consists in integrating over $q_0$ using the theorem of 
residues, and
then cutting the integral by $|{\bold q}|<\Lambda_3$. Since this procedure
is not Lorentz invariant, however, for quantities with two-body intermediate
states like
\refb{bubb} and \refb{fpi} it is necessary to specify the frame where the
integral is cut off, and then to 'boost' the result to a general
frame. If we cut off the integral in the frame ${\bold k}=0$, the
'boost' for Lorentz scalar quantities simply means to replace 
$k_0^2 \rightarrow k^2$ in the resulting
expressions. In this way we obtain
\begin{eqnarray}
1 + 2\frac{G}{M} \langle {\overline \psi} \psi \rangle_{3M} =  
1 - 24 G \int_{|{\footnotesize {\bold q}}|<\Lambda_3} 
\frac{d^3 q}{(2\pi)^3} \frac{1}{E_q} 
&=& \frac{m}{M} \label{gapr} \\ 
\Pi_{3M}(k) = 48 \int_{|{\footnotesize {\bold q}}|
<\Lambda_3} \frac{d^3 q}{(2\pi)^3}
\frac{E_q}{k^2-4 E_q^2+i\epsilon} &=& \int_{4 M^2}^{\Lambda^2} d\mu^2 
\frac{\rho(\mu^2)}{k^2-\mu^2+i\epsilon} \nonumber \\
\label{bubbr} \\
f_{\pi} = -12 g M \int_{|{\footnotesize {\bold q}}|<\Lambda_3} 
\frac{d^3 q}{(2\pi)^3}
\frac{1}{E_q\left(m_{\pi}^2-4 E_q^2\right)} &\equiv& - g M I(m_{\pi}^2), 
\label{fpir}
\end{eqnarray}
where $E_q=\sqrt{M^2+{\bold q}^2}$, and
the subscript 3M indicates that these quantities refer to the
3-momentum cut-off procedure. 
In the second equality of eq. \refb{bubbr} we introduced 
$\mu^2=4(M^2+{\bold q}^2)$ to express $\Pi_{}(k)$ as a dispersion integral
with ${\displaystyle \rho(\mu^2)= \frac{3 \mu^2}{4 \pi^2} 
\left(1-\frac{4M^2}{\mu^2}\right)^{\frac{1}{2}}}$, and 
$\Lambda^2=4(\Lambda_3^2+M^2)$. 

The following important relation between the condensate and the bubble
graph at $k=0$
\be
\langle {\overline \psi} \psi \rangle = M \Pi(0), 
\label{gene}
\ee
which can be derived from the chiral Ward identity for the pion decay vertex 
(see Appendix B), is satisfied in the 3M regularization scheme, i.e;
$\langle {\overline \psi} \psi \rangle_{3M} = M \Pi_{3M}(0)$. 
Therefore the gap equation \refb{gapr} and the pion
pole condition $1+2G\Pi(m_{\pi}^2)=0$ can be written as
\begin{eqnarray}
1+2 G \Pi_{3M}(0) &=& \frac{m}{M} \label{gapn} \\
\Pi_{3M}(m_{\pi}^2) - \Pi_{3M}(0) &=& m_{\pi}^2 I(m_{\pi}^2) = -\frac{m}{2GM},
\label{polen}
\end{eqnarray}
where $I(k^2)$ was introduced in \refb{fpir}. Eq. \refb{polen} shows
explicitly the Goldstone boson nature of the pion. The GOR relation is
then obtained as follows:
\begin{eqnarray}
-m \langle {\overline \psi} \psi \rangle_{3M} &=& 
\frac{Mm}{2G}\left(1-\frac{m}{M}\right) = -M^2m_{\pi}^2 I(m_{\pi}^2)
\left(1-\frac{m}{M}\right) \nonumber \\
&=& m_{\pi}^2 f_{\pi}^2 \left(1+m_{\pi}^2 C\right) \left(1-\frac{m}{M}\right) 
\label{gor}
\end{eqnarray}
with ${\displaystyle C=\frac{1}{I(m_{\pi}^2)} 
\frac{\partial I(m_{\pi}^2)}{\partial m_{\pi}^2}}$.   
In the last step we used the relation $f_{\pi}^2=-12 M^2 \\
\times I(m_{\pi}^2)/
(1+m_{\pi}^2 C)$, which follows from \refb{fpir} and 
${\displaystyle \frac{1}{g^2}=-\frac{\partial}{\partial k^2}\left(
k^2 I(k^2)\right)_{k^2=m_{\pi}^2}}$.

For quantities expressed in terms of LC variables, 
the 'invariant mass cut-off' introduced by
Lepage and Brodsky \cite{LB} is very convenient since it preserves all kinematic
symmetries. (We will denote this cut-off scheme by 'LB'.) 
It is applicable to Feynman diagrams containing 
intermediate states with
$n\geq 2$ particles propagating with respect to the LC 'time' variable $x^+$. 
For example, the bubble diagram ${\overline \Pi}(k,x)$  
involves a $q \overline{q}$ intermediate state with invariant mass squared
$2k_-\left(e_q+e_{k-q}\right)=
\left({\bold q}_{\perp}^2+M^2\right)/x(1-x)$. (We set ${\bold k}_{\perp}=0$.)
Requiring this to be smaller than
$\Lambda^2$ leads to the 'LB condition'
\be
{\rm LB}: \,\,\,\,\,\,\,\,\,\,
{\bold q}_{\perp}^2<\Lambda^2x(1-x)-M^2 \longrightarrow
x_1 \leq x \leq x_2,
\label{cond}
\ee
where 
${\displaystyle x_{1,2}=\frac{1}{2}\left(1 \pm \sqrt{1-4M^2/\Lambda^2}
\right)}$.
This prescription regulates the ultraviolett as well as the infrared
divergencies. The resulting regularized expressions 
${\overline \Pi}_{LB}(k,x)$, ${\overline \Pi}_{LB}(k)$ and $q_{v,LB}(x)$ 
are then given by
eqs. \refb{pibarx}, \refb{pibar} and \refb{result} with the condition 
\refb{cond}.

A crucial question of course is whether this LB regularization procedure is
consistent with the 3M procedure used above. For this, we
note that if we perform a variable 
transformation $x\rightarrow q_3$ with
\be 
x=\frac{1}{2}\left(1+\frac{q_3}{E_q}\right),\,\,\,\,\,\,\, 
(E_q=\sqrt{{\bold q}_{\perp}^2+q_3^2+M^2})
\label{trans}
\ee
in the expression for ${\overline \Pi}_{LB}(k)$, we exactly obtain
$\Pi_{3M}(k)$ of eq. \refb{bubbr} with $\Lambda^2=4(\Lambda_3^2+M^2)$. 
That is, the LB cut-off
$\Lambda$ is the same as the dispersion cut-off $\Lambda$ 
introduced in \refb{bubbr}. Similarly, if we evaluate  
the expression \refb{fpi} for $f_{\pi}$ in terms of LC coordinates,
\be
f_{\pi}=6 g M \int d x \int 
\frac{d^2q_{\perp}}{\left(2\pi\right)^3}
\frac{1}{{\bold q}_{\perp}^2+M^2-m_{\pi}^2x(1-x)},
\label{fpil}
\ee
impose the LB condition \refb{cond} and transform
$x\rightarrow q_3$ as above, we arrive at the expression eq. \refb{fpir}.
We therefore see that the 3M and LB cut-off procedures
are consistent in the sense that (i) they lead to the same expression
for $f_{\pi}$, and (ii) the quark distribution
regularized with the LB cut-off involves the quantity 
${\overline \Pi}_{LB}(k,x)$ (see eq. \refb{result}), the 
$x$-integral over which coincides with the
bubble graph $\Pi_{3M}(k)$, i.e; ${\overline \Pi}_{LB}(k)=\Pi_{3M}(k)$.  
   
At this point the question arises whether it is possible to
consistently regularize all integrals in terms of the LC coordinates
from the beginning, without making recourse to the regularization in terms
of the usual coordinates. For this purpose, the LB prescription has to be
extended, and this will be discussed in subsect. 3.5.


\section{NJL model on the light cone}
\setcounter{equation}{0}
In the ET quantization method used in the previous section, the introduction
of LC variables is merely a change of variables. In this section we will
investigate if we can obtain the same physical results in the framework
of LC quantization. For this purpose, the main task is first to derive
the form of the classical lagrangian in terms of the dynamical degrees
of freedom, from which the hamiltonian and the canonical quantization
conditions can be derived.

It is well known that in LC field theories not all components of the
fermion field $\psi$ are independent degrees of freedom \cite{BURK}. 
Namely, if we
introduce projection operators
\be
\Lambda_{\pm} = \frac{1}{2} \gamma^{\mp}\gamma^{\pm}
\label{proj}
\ee
and split the fermion field according to 
$\psi=\Lambda_+ \psi + \Lambda_- \psi \equiv \psi_+ + \psi_-$,
the component $\psi_-$ is not a dynamical degree of freedom since there
is no 'time' derivative $\partial_+$ acting on it in the lagrangian
(see eq. \refb{l2} below). Therefore, $\psi_-$ should be eliminated by
using the
equation of motion before the theory is quantized. This can be easily done
for theories which contain only fermionic bilinear terms, like free field
theories or Yukawa theories \cite{YAN}. 
For 4-fermi theories, however, the
'fermionic constraint equation' cannot be solved exactly \cite{REG,ITA}. 

In subsect.
3.1, we will use the $1/N$ expansion to derive the effective LC
lagrangian for $\psi_+$. 
Before plunging into the formalism, let us summarize here the result:
The lagrangian is expressed
in eq. \refb{end}, which shows that the
interaction term has the form obtained by naively using the
free field equations to eliminate $\psi_-$ (see eqs. \refb{free}), 
but there appears an
effective 4-fermi coupling constant ${\overline G}$ (see eq. \refb{geff}) 
instead of the
original $G$. In subsect. 3.2 we discuss the Feynman rules
which follow from this effective lagrangian, and in subsects. 3.3 and 3.4
we re-derive the ${\overline q} q$ t-matrix in the pionic channel, the
pion decay constant and the quark LC momentum distribution in the pion.
There we will show that the results for these quantities are formally the
same as obtained in the ET theory. In subsect. 3.5 we discuss two possible
regularization procedures in terms of LC coordinates: An extended version
of the LB scheme, and an alternative method \cite{TRAN} which avoids the 
introduction
of a cut-off for the longitudinal momentum components. 

\subsection{Effective LC lagrangian in the 1/N expansion}

To define the $1/N$ expansion, we write as usual 
${\displaystyle G=\frac{G_0}{N}}$ with $G_0$ independent of $N$.
Let us first discuss the ${\cal O}(N)$ of the terms in the lagrangian
\refb{lag}. Splitting ${\overline \psi}\psi=<{\overline \psi}\psi>
+\left(:{\overline \psi}\psi:\right)$, where $<...>$ denotes a contraction
with respect to the fermi fields, we have $<{\overline \psi}\psi>\propto N$
and $\left(:{\overline \psi}\psi:\right)\propto \sqrt{N}$.
\footnote{This follows from $<\left({:\overline \psi}\psi:\right)^2>\propto
N$. A similar consideration shows that in the decomposition
$\left(:{\overline \psi}\psi:\right)^2 = 
<\left(:{\overline \psi}\psi:\right)^2> + 
\left(:\left(:{\overline \psi}\psi:\right)
^2:\right)$ {\em both} terms are of ${\cal O}(N)$.}
Therefore the lagrangian \refb{lag} contains a c-number part of ${\cal O}(N)$,
a 1-body part of order ${\cal O}(\sqrt{N})$, and
a 2-body part of order ${\cal O}(1)$. Our aim here is to give a similar
decomposition of the lagrangian in the LC theory, i.e; to derive the
1-body part to ${\cal O}(\sqrt{N})$ and the 2-body part to ${\cal O}(1)$.
 
To formulate the $1/N$ expansion in the LC theory, it is convenient to 
introduce bosonic auxiliary fields $\sigma$ and ${\bold \pi}$. 
The lagrangian \refb{lag} is equivalent to
\be 
\Lag= \overline{\psi}(i\fslash{\partial} - \hat{M} )\psi
       -\frac{G_0}{N}\left(\sigma^2+{\bold \pi}^2\right),
\label{l1}
\ee  
where
\be
\hat{M}=m+2 \frac{G_0}{N}\left(\sigma + i {\bold \pi} \cdot {\bold \tau}
\gamma_5\right).
\label{hat}
\ee
We then rewrite this lagrangian in terms of LC variables, making use of
the projection operators \refb{proj}. Some useful relations for quark
bilinears are given in Appendix A, and we obtain
\begin{eqnarray} 
\Lag &=& \sqrt{2} \left(\psi_+^{\dagger} i \partial_+ \psi_+ + 
\psi_-^{\dagger}  i \partial_- \psi_-\right) 
-\frac{G_0}{N}
\left(\sigma^2+{\bold \pi}^2\right) \nonumber  \\
&-& \psi_+^{\dagger}
\left( i {\bold \alpha}_{\perp}\cdot {\bold \partial}_{\perp}
+ \gamma^0 \hat{M}\right)\psi_-  - \psi_-^{\dagger}
\left( i {\bold \alpha}_{\perp}\cdot {\bold \partial}_{\perp}
+ \gamma^0 \hat{M}\right)\psi_+ . \label{l2}
\end{eqnarray}
Since we introduced the bosonic auxiliary fields, the fermionic
constraint equation 
${\displaystyle \partial \Lag / \partial \psi_-^{\dagger} = 0}$ can be
solved explicitly: \footnote{We use the following definition of 
${\displaystyle \frac{1}{\partial_-}}$:
\begin{eqnarray}
\left(\frac{1}{\partial_-} f\right)(x^-) 
\equiv \frac{1}{2} \int_{-\infty}^{\infty} d y^-
\epsilon(x^- - y^-) f(y^-) \nonumber 
\end{eqnarray}
with $\epsilon(x)=\Theta(x)-\Theta(-x)$.} 
\be
\psi_-=\frac{1}{\sqrt{2} i \partial_-} \left(i {\bold \alpha}_{\perp}\cdot 
{\bold \partial}_{\perp} + \gamma^0 \hat{M}\right)\psi_+,
\label{psim}
\ee
and inserting this into eq. \refb{l2} we obtain
\begin{eqnarray}
\Lag &=& \sqrt{2} \psi_+^{\dagger} i \partial_+ \psi_+
-\frac{G_0}{N} \left(\sigma^2+{\bold \pi}^2\right) \nonumber \\ 
& - & \frac{1}{\sqrt{2}} \psi_+^{\dagger}
\left( i {\bold \alpha}_{\perp}\cdot {\bold \partial}_{\perp}
+ \gamma^0 \hat{M}\right) \frac{1}{i \partial_-} 
\left( i {\bold \alpha}_{\perp}\cdot {\bold \partial}_{\perp}
+ \gamma^0 \hat{M}\right) \psi_+ .
\label{l3}
\end{eqnarray}

We now separate the leading order ($N$) of the 'c-number part'
of the $\sigma$ field by writing $\sigma=\sigma_0 N + \hat{\sigma}$.
Here $\hat{\sigma}$ contains the leading order ($\sqrt{N}$) of the 
'fluctuation part', as well as the next-to-leading order ($N^0$) of
both the c-number and the fluctuation part. From the form
of the lagrangian \refb{l3} it is clear that up to the order which
concerns us here (that is, ${\cal O}(\sqrt{N})$ for the 1-body part,
${\cal O}(N^0)$ for the 2-body part) it is sufficient to consider
$\hat{\sigma}$ (and also ${\bold \pi}$) as the fluctuation piece of
${\cal O}(\sqrt{N})$. We define the constituent quark mass as 
\be
M= m + 2 G_0 \sigma_0.
\label{cons}
\ee  
To split the lagrangian into 'free' and 'interaction' parts, it is
convenient to follow ref. \cite{YAN} and define the fermion field
\be
\Psi\equiv \Psi_+ + \Psi_-
\label{split}
\ee
with
\be
\Psi_+ \equiv \psi_+; \,\,\,\,\,\,\,\,\,\,\,\,\,\,
\Psi_- \equiv \frac{1}{\sqrt{2} i \partial_-} 
\left(i {\bold \alpha}_{\perp}\cdot 
{\bold \partial}_{\perp} + \gamma^0 M \right)\psi_+.
\label{free}
\ee
In terms of this field the lagrangian \refb{l3} can be written as
\begin{eqnarray}
\Lag &=& \Lag_0 - 2\,\, \frac{G_0}{N}\,\, {\overline \Psi} 
\left(\hat{\sigma} + i {\bold \pi} \cdot {\bold \tau} \gamma_5\right) \Psi
-\frac{G_0}{N} \left(\hat{\sigma}^2+{\bold \pi}^2\right) - 
2 G_0 \sigma_0 \hat{\sigma}
\nonumber \\
&-& 2 \left(\frac{G_0}{N}\right)^2 {\overline \Psi} 
\left(\hat{\sigma} + i {\bold \pi} \cdot {\bold \tau} \gamma_5\right)
\gamma^+ \frac{1}{i \partial_-}\left(\hat{\sigma} + i {\bold \pi} \cdot 
{\bold \tau} \gamma_5\right) \Psi 
\label{l4}
\end{eqnarray}
with
\begin{eqnarray}
\Lag_0 &=& {\overline \Psi}\left( i\fslash{\partial} - M \right) \Psi 
\nonumber \\
&=& \sqrt{2} \psi_+^{\dagger} i \partial_+ \psi_+ - 
\frac{1}{\sqrt{2}} \psi_+^{\dagger}
\left( i {\bold \alpha}_{\perp}\cdot {\bold \partial}_{\perp}
+ \gamma^0 M \right) \frac{1}{i \partial_-} 
\left( i {\bold \alpha}_{\perp}\cdot {\bold \partial}_{\perp}
+ \gamma^0 M \right) \psi_+. \nonumber \\
\label{lag0}
\end{eqnarray}
We note that up to the required order it is sufficient to replace the
two fermi fields in the last term of \refb{l4} by a contraction.
To exhibit the dependence of the lagrangian on the bosonic auxiliary
fields more explicitly, we express it in terms
of the combined field $\Phi \equiv (\hat{\sigma}, {\bold \pi})$
(i.e; $\Phi^{(0)}=\hat{\sigma},\,\,\,\Phi^{(i)}=\pi^i$) and the corresponding
vertex operator $\Gamma\equiv (1,i\gamma_5 {\bold \tau})$ as
\be
\Lag=\Lag_0  - \frac{1}{2} \Phi^{(\alpha)} \vec{A}_{\alpha \beta}
\Phi^{(\beta)} - \Phi^{(\alpha)} B_{\alpha}, 
\label{l5}
\ee
where we introduced 
\begin{eqnarray}
\stackrel{\rightarrow}{A}_{\alpha \beta} &=& 
2 \frac{G_0}{N} \delta_{\alpha \beta} +
\left(2 \frac{G_0}{N}\right)^2 \langle {\overline \Psi} \Gamma_{\alpha}
\gamma^+ \frac{1}{i \stackrel{\rightarrow}{\partial}_-} \Gamma_{\beta} \Psi
\rangle 
\label{aa} \\
B_{\alpha} &=& 2 \delta_{\alpha 0} G_0 \sigma_0 + 
2 \,\,\frac{G_0}{N}\,\, {\overline \Psi} 
\Gamma_{\alpha} \Psi. 
\label{bb}
\end{eqnarray}
We now eliminate the auxiliary boson fields using 
$\partial \Lag / \partial \Phi = -A \Phi - B = 0$, 
which gives formally
\be
\Phi = - A^{-1} B; \,\,\,\,\,\,\,\,\,\,\,\,\,\,\,
\Lag=\Lag_0 + \frac{1}{2} B^T A^{-1} B.
\label{sol} 
\ee
The matrix operator $A$ introduced above is defined by
\footnote{The quantity $\stackrel{\leftarrow}{A}$ involves
$\frac{1}{\stackrel{\rightarrow}{\partial}_-} 
= \frac{1}{\partial_-}$ acting to the right, while
$\stackrel{\leftarrow}{A}$ involves 
$\frac{-1}{\stackrel{\leftarrow}{\partial}_-}$ acting to the left. The
symmetrized form of $A$ in \refb{sym} appears naturally if we take the
functional derivative of the action $\int d^4 x \Lag(x)$ with respect to 
$\Phi(x)$, using the definition of $1/\partial_-$. We note that also in 
the lagrangian eq. \refb{l5}, 
$\stackrel{\rightarrow}{A}_{\alpha \beta}$ can be replaced by 
$A_{\alpha \beta}$, since
this leaves the action unchanged.}
\be
A_{\alpha \beta} \Phi^{(\beta)} \equiv \frac{1}{2} \left(\stackrel
{\rightarrow}{A}_{\alpha
\beta} \Phi^{(\beta)} + \Phi^{(\beta)} 
\stackrel{\leftarrow}{A}_{\beta \alpha}\right).
\label{sym}      
\ee
Due to parity
and/or isospin invariance, $A_{\alpha \beta}\propto \delta_{\alpha \beta}$, 
and we therefore define
\be
A_{\alpha \beta} \equiv \delta_{\alpha \beta} \,\, 
2\,\,\frac{G_0}{N}\,\, \left(1+2\,\,\frac{G_0}{N}\,\,F\right),
\label{defa}
\ee
where the quantity 
\be
F = \frac{1}{2}\left(<{\overline \Psi} \gamma^+ \frac{1}{i 
\stackrel{\rightarrow}{\partial}_-} 
\Psi>+<{\overline \Psi} \gamma^+ \frac{-1}
{i \stackrel{\leftarrow}{\partial}_-} \Psi>
\right)
\label{deff}
\ee
is of ${\cal O}(N)$. 
 
$\sigma_0$ is determined by the requirement
$<\Phi>=-A^{-1} <B>\equiv 0$. Due to \refb{bb} and \refb{cons}, this 
requirement $<B>=0$ leads to the gap equation
\be
M=m-2G <{\overline \Psi} \Psi>. 
\label{galc}
\ee
Since in this leading order ($N$)
we have $<{\overline \Psi} \Psi>=<{\overline \psi} \psi>$,
this is formally the same gap equation as obtained in the ET theory,
eq. \refb{gap}.
 
Since $<B>=0$, the interaction term in the lagrangian
\refb{sol} becomes 
${\displaystyle \frac{1}{2} \left(:B:\right) A^{-1} \\ \left(:B:\right)}$.   
We therefore arrive at the final form of the LC lagrangian: 
\footnote{Our notation used in eq. \refb{end}
anticipates the fact derived in the next subsection that in coordinate
space the quantity ${\overline G}$  
$\propto \delta(x^- -y^-)$, i.e; that the interaction lagrangian is local
although $F$ of eq. \refb{deff} is apparently non-local.} 
\be
\Lag = \Lag_0 + {\overline G}
\left( \left(:{\overline \Psi} \Psi:\right)^2 -  
\left(:{\overline \Psi} \gamma_5 {\bold \tau} \Psi:\right)^2 \right),
\label{end}
\ee
where $\Lag_0$ of eq. \refb{lag0} contains the 1-body part of 
${\cal O}(\sqrt{N})$,   
and the 2-body part
is of order $N^0$ and involves the effective coupling constant
\be
{\overline G} = \frac{G}{1+2\,\,G\,\,F}\,,
\label{geff}
\ee
with $G=G_0/N$.
One should note that the lagrangian \refb{end} contains only $\psi_+$ as a
dynamical component, since we {\em defined} the field $\Psi$ in terms
of $\psi_+$ in eqs. \refb{split}, $\,$ \refb{free}.  
Our result is that {\em the effective interaction lagrangian in the LC theory  
up to ${\cal O}(N^0)$ can be obtained from eq. \refb{lag} by using the 
free field equations
to eliminate $\psi_-$, but the coupling constant $G$ is replaced by
the effective coupling constant ${\overline G}$ of eq. \refb{geff}}.
(The explicit expression for the quantity $F$ will be given in eq.
\refb{exf} below.) 

Once we have derived the classical lagrangian for $\psi_+$, the 
quantization proceeds as usual by imposing anti-commutation
relations between $\psi_+$ and $\Pi_+=\sqrt{2}i\psi_+^{\dagger}$ at
equal LC times $x^+$. The mode expansion of the free field reads
\footnote{Properties of the LC spinors $u_{\pm}=\Lambda_{\pm} u$ and
$v_{\pm}=\Lambda_{\pm} v$ are summarized in Appendix A.}
\be
\psi_+(x)=\int \frac{d^3p}{\left(2\pi\right)^{3/2}}\frac{1}{\sqrt{2 p_-}}
\sum_s \left[b_{{\footnotesize {\bold p}},s} u_+({\bold p},s) 
e^{-i{\overline p}\cdot x}
+ d^{\dagger}_{{\footnotesize {\bold p}},s} v_+({\bold p},s) 
e^{i{\overline p}\cdot x}\right]
\label{mod}
\ee
with $d^3 p=dp_-d^2p_{\perp}\,\,\,$, ${\bold p}=(p_-, {\bold p}_{\perp})$,
$p \cdot x \equiv p_+ x^+ + p_- x^- - {\bold p}_{\perp}\cdot 
{\bold x}_{\perp}$, and ${\overline p}^{\mu}$ was defined below \refb{sbar}. 
Using the mode expansion given above, one obtains for the free fermion 
propagator 
\begin{eqnarray}
{\overline S}_{++}(p)&\equiv& F.T. \left(-i<0|{\tilde T}\left(\psi_+ \psi_
+^{\dagger}\right)|0>\right) \nonumber \\
&=& \frac{\Lambda_+ \sqrt{2}\,\,p_-}{p^2-M^2+i\epsilon}, 
\label{pp}
\end{eqnarray}
where ${\tilde T}$ orders the operators with respect to $x^+$. 
The quantity $1/(p^2-M^2+i\epsilon)$ has been given in terms of LC variables 
in eq. \refb{den}.


\subsection{Feynman rules, gap equation and effective coupling constant}

To calculate a fermionic Green function
\be
<0|{\tilde T}\left(\psi_+(x_1)  \ldots \psi_+^{\dagger}(x_n)\right)|0>,
\label{g1}
\ee
one could formulate the Feynman rules 
directly in terms of the field $\psi_+$, using the lagrangian \refb{end} and
the definition of $\Psi$ in terms of $\psi_+$ given in eq. \refb{split}. 
In this method, one uses
the propagator eq. \refb{pp} for internal fermion lines, and 4-fermi 
vertices
of the following form, referring e.g; to the pionic $q {\overline q}$ channel,  
\begin{eqnarray}
\lefteqn{-2i{\overline G} X_k^i({\bold q'}) \, X_k^i({\bold q})} \nonumber \\
& \equiv &  
 -2i{\overline G} \left[\gamma_5 \gamma^0 \tau_i \left(Z({\bold q'})-
Z({\bold q'}-{\bold k})
\right)\right] \,\,\left[\gamma_5 \gamma^0 \tau_i \left(
Z({\bold q}-{\bold k})-Z({\bold q})
\right)\right], \nonumber \\ 
\label{vert} 
\end{eqnarray}
where the external momenta are assigned as in fig. \ref{mom}, and 
\be
Z({\bold p})=\left({\bold \alpha}_{\perp}\cdot {\bold p}
_{\perp} + \gamma^0 M\right)\frac{1}{\sqrt{2} \,\,p_-}.
\label{defz}
\ee
A much more convenient method, however,
is first to calculate the Green function
\be
<0|{\tilde T}\left(\Psi(x_1)  \ldots \Psi^{\dagger}(x_n)\right)|0>,
\label{g2}
\ee
and then to project out \refb{g1} by applying the
projection operator $\Lambda_+$ to each external fermion leg. In this method,
which will be used in the next section to obtain the $q {\overline q}$
t-matrix in the pionic channel, one uses the propagator
\begin{eqnarray}
{\overline S}(p)&\equiv& F.T. \left(-i<0|{\tilde T}\left(\Psi \Psi^{\dagger}
\right)|0>\right) = \frac{\overline {\fslash{p}}+M}{p^2-M^2+i\epsilon}
\gamma^0 = S(p) - \frac{\gamma^+ \gamma^0}{2 p_-} \nonumber \\
\label{sbar1}
\end{eqnarray}
for each internal fermion line, and (constant) 4-fermi vertices which
differ from those in the ET treatment of the NJL model
only by the replacement $G\rightarrow {\overline G}$. 
The form of ${\overline S}(p)$ given in eq. \refb{sbar1} can be easily
derived as follows \cite{YAN}:
Splitting ${\overline S}(p)$ into the four components ${\overline S}_{ij}\,\,\,
(i,j=+,-)$, where ${\overline S}_{++}$ is given by eq. \refb{pp}, we have, 
using the form of $\Psi$ as given by eq.\refb{split} and the Feynman
propagator \refb{fey} (see also Appendix A): 
\begin{eqnarray}
{\overline S}_{+-}(p)&=&{\overline S}_{++}(p)Z({\bold p}) 
= \Lambda_+ S(p) \Lambda_-
\nonumber \\
{\overline S}_{-+}(p)&=&Z({\bold p}){\overline S}_{++}(p) 
= \Lambda_- S(p) \Lambda_+
\nonumber \\  
{\overline S}_{--}(p)&=&Z({\bold p}){\overline S}_{++}(p) Z({\bold p}) = 
\Lambda_- S(p) \Lambda_- - \frac{\Lambda_-}{\sqrt{2}\,\, p_-}.
\label{ide1}
\end{eqnarray}
>From these relations one easily obtains eq.\refb{sbar1}.
This propagator ${\overline S}(p)$ has been introduced already in
eq. \refb{sbar} by separating the contact term from the Feynman
propagator $S(p)$.   

Clearly, if the
ET and the LC treatments of the NJL model are equivalent, the contributions
due to additional term $-\gamma^+ \gamma^0/2p_-$ in the propagator
and the modifications due to the replacement $G\rightarrow {\overline G}$
in the vertices
should 'cancel' for physical observables. These are the same kinds of
cancellations as observed for the $\sigma$-model in ref. \cite{YAN}.

Using the propagator \refb{sbar1}, it is now easy to derive the explicit forms
of the gap equation and the quantity F, which determines the effective
coupling constant (see eqs. \refb{deff} and \refb{geff}). 
For the gap equation we obtain from \refb{cons} and \refb{galc}
\be
M = m - 2 G \langle {\overline \Psi} \Psi \rangle = 
m + 12iG\kint{q}tr_D\left({\overline S}(q)\gamma^0\right), 
\label{gapl}
\ee
and since from eq. \refb{sbar1} we have 
$tr_D\left({\overline S}(q)\gamma^0\right)=
tr_D\left(S(q)\gamma^0\right)$, we see that the gap equation is formally
the same
as in the ET theory. Using \refb{den} and performing the $q_+$ integral 
we obtain
\be
M-m= 48 i G M \kint{q} \frac{1}{q^2-M^2+i\epsilon} = 
24 G M \int_0^{\infty} \frac{dq_-}{q_-} \int \frac{d^2 q_{\perp}}
{\left(2 \pi\right)^3}.
\label{ga}
\ee
Although it seems from this expression that in the case $m=0$ only the
trivial solution ($M=0$) exists, we have to note that 
the theory has not yet been regularized. (See subsect. 3.5.)

Let us now turn to the quantity $F$ of eq.\refb{deff}. 
Using the definition of 
${\displaystyle \frac{1}{\partial_-}}$ given earlier, we have for the 
quantity $A_{\alpha \beta}\equiv \delta_{\alpha \beta} a$ of eq.\refb{defa}
\be
a(z^-)= 2\,\,G\,\,\left[ \delta(z^-) + 3 G\,  \epsilon(z^-) \, tr_D
\left({\overline S}(z^-) \gamma^0 \gamma^+ 
- {\overline S}(-z^-) \gamma^0 \gamma^+\right)\right]
\ee
where $z^-=x^--y^-$. The Fourier transform of this is
\be
a(k_-)= 2\,\,G\,\,\left(1  - 6i G \int \frac{d^4 q}{\left(2\pi\right)^4}
tr_D \left({\overline S}(q) \gamma^0 \gamma^+\right)
\left(\frac{{{\cal P}}}{q_-+k_-} +\frac{{{\cal P}}}{q_--k_-}\right) \right).
\label{af}
\ee   
Again we note that in this expression ${\overline S}(p)$ can be replaced
by the Feynman propagator $S(p)$ without any modification. 
Performing the
$q_+$ integral using \refb{den} and introducing $q_-=k_-x$ 
we see that $a$ (or $F$) is actually a constant, and so is 
${\overline G}$ of eq.\refb{geff}.
The expression for $F$ becomes
\be
F= -6 {\cal P} \int_0^{\infty} dx \left(\frac{1}{x+1} 
+\frac{1}{x-1}\right) \int \frac{d^2 q_{\perp}}
{\left(2 \pi\right)^3}.
\label{exf} 
\ee

\subsection{Pionic t-matrix}

According to the discussion at the beginning of the previous subsection,
to obtain the $q {\overline q}$ t-matrix in the LC formulation,
we first determine the t-matrix associated with the amputated connected
Green function $<0|\tilde{T}\left(\Psi \Psi \Psi^{\dagger} \Psi^{\dagger}
\right)|0>$. In the ladder approximation this is obtained by summing
the bubble
graphs ${\overline \Pi}(k)$, which have been introduced already in sect. 2, 
in the same way as in the ET theory as
\be
{\overline t}(k) = \left(\gamma^0 \gamma_5 \tau_i \right)
{\overline \tau}(k) \left(\gamma^0 \gamma_5 \tau_i \right)
\label{tb}
\ee
with
\begin{eqnarray}
{\overline \tau}(k)& = &\frac{-2 i {\overline G}}{1+2 {\overline G}\,\,\,
{\overline \Pi}(k)} \label{taub} \\
{\overline \Pi}(k) &=& 6i \kint{q} tr_D \left(\gamma^0 \gamma_5
{\overline S}(q)\gamma^0 \gamma_5 {\overline S}(q-k)\right).
\label{pib}
\end{eqnarray}
The t-matrix associated with the amputated connected
Green function \\ $<0|\tilde{T}\left(\psi_+ \psi_+ \psi_+^{\dagger} \psi_+
^{\dagger}\right)|0>$ is then simply obtained by projection (see fig.
\ref{mom} for the assignment of the external momenta):
\begin{eqnarray}
& & {\overline S}_{++}(q'){\overline S}_{++}(q'-k)\,\, {\tilde t} \,\,
{\overline S}_{++}(q-k){\overline S}_{++}(q) \nonumber \\
&=& \Lambda_+ \left({\overline S}(q') \gamma^0 \gamma_5 \tau_i
{\overline S}(q'-k) \right)\Lambda_+
\,\,{\overline \tau}(k)\,\,\Lambda_+ \left({\overline S}(q-k) \gamma^0 
\gamma_5 \tau_i {\overline S}(q) \right)\Lambda_+ \nonumber \\
\label{pro}
\end{eqnarray}
The result of this projection is 
${\tilde t} = X_k^i ({\bold q'})\, {\overline \tau}(k)\, X_k^i ({\bold q})$, 
i.e; the vertices
$\left(\gamma^0 \gamma_5 \tau_i \right)$ are replaced by the momentum dependent
vertices $X_k^i$ of eq.\refb{vert}. Of course, the pole structure is not
affected by this projection.

Using eq.\refb{geff}, we can rewrite the reduced t-matrix \refb{taub} as
\be
{\overline \tau}(k) = \frac{-2 i G}{1+2 G \left({\overline \Pi}(k)+F\right)} 
\ee
Comparing with eq. \refb{taupi}, we see that we should have the relation 
\be
\Pi(k) = {\overline \Pi}(k)+F
\ee
with $F$ given by \refb{exf},
in order that the reduced t-matrices, and in particular the pole parts,
in ET and LC theories are the same. On the other hand, by explicit
evaluation we found that $\Pi(k) = {\overline \Pi}(k)+R$ (see eq.\refb{piece}), 
where $R$ is given by \refb{rx2}. Formally, the relation
\be
F=R
\label{equal}
\ee
holds, but to establish it from \refb{rx2} and \refb{exf} one has to
introduce variable shifts in divergent integrals.

The Goldstone nature of the pion in the present LC treatment follows
by noting that the relation \refb{gene}, which now reads 
\be
<{\overline \Psi} \Psi> = M \left({\overline \Pi}(0)+R\right),
\label{valid}
\ee
is formally satisfied. The
explicit verification by using \refb{ga}, \refb{pibarx} and \refb{rx2}, 
however, again
involves variable shifts in divergent integrals. 
If one nevertheless assumes the validity of
\refb{valid} and \refb{equal},
one can write the
gap equation \refb{gapl} and the pion pole condition $1+2{\overline G} \,\, 
{\overline \Pi}(m_{\pi}^2)=0$ as follows:
\begin{eqnarray}
1+2{\overline G}\,\, {\overline \Pi}(0) &=& \frac{\overline m}{M} 
\label{gaa} \\
{\overline \Pi}(m_{\pi}^2) - {\overline \Pi}(0) &=& m_{\pi}^2 
I(m_{\pi}^2) = -\frac{{\overline m}}{2{\overline G}M},
\label{polen1}
\end{eqnarray}
where $I(k^2)$ was introduced in \refb{fpir}. Here we used \refb{geff} and
\refb{gene} to express the gap equation in terms of ${\overline G}$ and
${\overline \Pi}(0)$, and introduced 
\be
\overline m = \frac{m}{1+2GF} = m \frac{\overline G}{G}.
\label{meff}
\ee
Eq. \refb{polen1} shows the Goldstone nature of the pion. We note the
the GOR relation is satisfied with the current quark mass $m$ (not
${\overline m}$). Its derivation follows the same steps as outlined
in eq. \refb{gor}.

\subsection{Bound state matrix elements and pionic LC wave function}

In contrast to the BS wave function introduced in sect.2, a 'LC
wave function' usually refers to the solution of a Schr\"odinger-type
(or, more precisely, TDA-type) equation 
which can be derived
directly in the hamiltonian formalism (see Appendix C). Here we want to show
how this kind of wave function can also be obtained by using the projection
methods introduced in the previous sections.

In the same way as in the ET theory, we first define a BS wave function
${\overline \Phi}$
in the LC theory by
\be
{\overline S}(q'){\overline S}(q'-k)\,\, {\overline t} \,\,
{\overline S}(q-k){\overline S}(q)
\rightarrow
\frac{i}{2 k_-} \frac{\overline {\Phi}_k^i(q') \overline{\Phi}^{i \dagger}
_k(q)}{k_+-e_k+i\epsilon}
\,\,\,\,\,\,\,\,\,{\rm as}\,\,k_+\rightarrow e_k.
\ee
This wave function ${\overline \Phi_k(q)}$ is given by \refb{wave} with 
$S \rightarrow {\overline S}$.
The BS wave function for the dynamical (+) fermion components 
($\tilde{\Phi}$) is defined by (see eq. \refb{pro})
\be
{\overline S}_{++}(q'){\overline S}_{++}(q'-k)\,\, {\tilde t} \,\,
{\overline S}_{++}(q-k){\overline S}_{++}(q)
\rightarrow
\frac{i}{2 k_-} \frac{\tilde {\Phi}_k^i(q') \tilde {\Phi}^{i \dagger}
_k(q)}{k_+-e_k+i\epsilon}
\,\,\,\,\,\,\,\,\,{\rm as}\,\,k_+\rightarrow e_k.
\label{dyn}
\ee
It is clear that
\be
\tilde{\Phi}^i_k(q) = \Lambda_+ \overline{\Phi}_k(q) \Lambda_+ = 
g \biggl(\Lambda_+ \overline{S}(q) \gamma^0 \gamma_5 
\tau_i \overline{S}(q-k) \Lambda_+ \biggr) 
\label{wave1}
\ee
Due to $\Lambda_+\gamma^+=\gamma^+\gamma^0\Lambda_+=0$ and \refb{sbar1},
the propagator $\overline{S}$ can be replaced by the Feynman propagator
$S$ in this expression, and $\tilde{\Phi}$ agrees with the $++$
component of the BS wave function \refb{wave}.  

The calculation of bound state matrix elements in the LC theory can then
proceed by using these BS wave functions in the same way as in the ET
theory. As in the case of the Green functions, one has two choices:
One is to use the wave function $\tilde{\Phi}$
together with the vertices and
propagators defined in terms of the field $\psi_+$, and the other, more
convenient method
is to use the wave function ${\overline \Phi}$ together with the 
vertices and propagators
defined for the field $\Psi$. 
In Appendix D it is shown that the calculation of the pion decay constant
and the valence quark distribution in the pion along
these lines leads to the same results as in the ET theory, namely
eq.\refb{fpi} and eq. \refb{qv}.  

Another method to calculate matrix elements involving bound state like those
in \refb{dfpi} and \refb{qv1}, which is more specific to LC theories, 
is to use the
Fock space expansion of the pion state $|ki>$. Let us explain this
approach here in order to make contact to the concept of Schr\"odinger-type
LC wave functions. It is well
known \cite{BER} that, for interactions which do not depend on the 
'time' component
of the relative momentum ($q_+$ here), the Schr\"odinger-type (equal 'time') 
wave function is obtained by integrating the BS wave function over
$q_+$ :
\begin{eqnarray}
& &{\Phi}^{i(S)}_k({\bold q}) = \int \frac{d q_+}{2\pi i} 
\tilde{\Phi}^i_k(q) \nonumber \\
&=& g \int \frac{dq_+}{2\pi i} \left(\Lambda_+ {\overline S}(q) 
\gamma^0 \gamma^5 
\tau_i {\overline S}(q-k) \Lambda_+\right).
\label{sch}
\end{eqnarray}   
As shown in Appendix C, this 'LC wave function' satisfies the same 
Schr\"odinger-type 
equation as derived directly in the hamiltonian formalism. 
If we use the propagator \refb{sbar1}
with the pole part given in
eq.\refb{den} in this expression and consider the case $k_->0$, 
we get the restrictions
$q_->0$ and $q_--k_-<0$, and therefore the LC wave function \refb{sch} 
is non-zero only for $0<x<1$, where $x=q_-/k_-$. 
To go from the Dirac to the spin representation,
we multiply eq.\refb{sch} from left by 
$\left(\sqrt{2}q_-\right)^{-1/2} 
u_+^{\dagger}({\bold q},s_1)$ and from right by
$\left(\sqrt{2}(k_--q_-)\right)^{-1/2} v_+({\bold k}-{\bold q},s_2)$.
Multiplying also the corresponding isospin ($t$) and color ($c$) 
wave functions,
and denoting $a\equiv(s,t,c)$, we obtain after a simple calculation
\begin{eqnarray}
\Phi^{i(S)}_{k,a_1a_2}({\bold q})&=&\frac{g}{2 k_-} 
\frac{1}{\sqrt{x(1-x)}} \frac{1}{m_{\pi}^2x(1-x)-{\bold q}_{\perp}^2-M^2}
\left(\chi^{\dagger}_{t_1} \tau_i \chi_{t_2}\right) \delta_{c1,c2} 
\nonumber \\
&\times & u_+^{\dagger}({\bold q},s_1)\gamma_5 \left(
{\bold \gamma}_{\perp}\cdot {\bold q}_{\perp}+M\right)
v_+({\bold k}-{\bold q},s_2).
\label{final}
\end{eqnarray}
The conventional normalization
\be
\int \frac{d^2 q_{\perp}}{\left(2 \pi\right)^3} \int_0^1 dx
\sum_{a_1a_2}
|\Phi^{i(S)}_{k,a_1a_2}(q_-=k_-x, {\bold q}_{\perp})|^2
=1
\ee
can be checked from eq. \refb{final}, using $1/g=
\left[-\partial {\overline \Pi}/{\partial k^2}\right]^{\frac{1}{2}}$ 
($k^2=m_{\pi}^2$)
with ${\overline \Pi}(k)$ given by \refb{pibar}.
The Fock space expansion of the pionic bound state in the TDA is then given by
\be
|ki>=\sqrt{2} \int_0^{k_-}d q_- \int d^2q_{\perp} 
\sum_{a_1 a_2} \Phi^{i(S)}_{k,a_1a_2}
({\bold q}) b^{\dagger}_{{\footnotesize {\bold q}}, a_1} 
d^{\dagger}_{{\footnotesize {\bold k}} - {\footnotesize {\bold q}},a_2}|0>.
\label{fock}
\ee
If we note that ${\overline \psi}(0) \gamma^+ \psi(z^-)=
\sqrt{2}\psi_+^{\dagger}(0) \psi_+(z^-)$ in eq.\refb{qv1}, we can
use the mode expansion \refb{mod} (supplemented by the isospin and color
parts) and the Fock expansion \refb{fock}
to obtain after a simple calculation
\be
q_v(x)=\int\frac{d^2 q_{\perp}}{\left(2\pi\right)^3} \sum_{a_1a_2}
|\Phi^{i(S)}_{k,a_1a_2}(q_-=k_-x, {\bold q}_{\perp})|^2.
\label{spins}
\ee
The evaluation of the sum then leads to the same result as obtained
in the ET theory, namely eq.\refb{result}. 

In a similar way, by noting that the $+$ component of the axial vector
current is given by
${\displaystyle \frac{1}{\sqrt{2}}\psi_+^{\dagger} \gamma_5 {\bold \tau}
\psi_+}$, we obtain from eq. \refb{dfpi} the expression \refb{fpil} (yet 
unregularized).

\subsection{Regularization schemes in LC theory}

In the previous subsections we have shown that the NJL model on the LC
gives formally the same results for the pion decay constant, the gap
equation, the pion pole condition and the quark distribution in the
pion as the ET theory. In sect. 2 we noted that the bubble graphs
in the LC theory (like ${\overline \Pi}(k,x)$ and ${\overline \Pi}(k)$)
can be regularized in a covariant way consistently with the 3M 
regularization scheme in the ET theory.
To give consistent regularization prescriptions for
tadpole graphs in the LC theory, like the quark condensate or the quantity 
$F$ of eq. \refb{exf},
turns out to be a very difficult task which we do not attempt to solve
explicitly in this work. We will instead make use of the 
general relation \refb{valid}. We have shown that this relation holds
formally in the LC theory, if one performs variable shifts in divergent
longitudinal momentum ($q_-$) integrals. Here we will assume that 
\refb{valid} and also \refb{equal} are valid 
even in the regularized theory. This method of 
{\em defining}
the quark condensate in LC theories in terms of correlation
functions has been used in several recent works \cite{DEF}.

Following this method, we will give two possible regularization schemes
below. The first one ('extended LB scheme') is
equivalent to the 3M cut-off scheme in the ET formulation. The second
one ('transverse cut-off scheme'), which was proposed in ref. \cite{TRAN}, 
avoids the introduction of a cut-off
for the longitudinal momentum component.

\subsubsection{Extended LB scheme}

In addition to the LB condition \refb{cond} for the 2-point functions, 
in the 'extended LB scheme' one assumes the validity of 
eqs. \refb{valid} and \refb{equal}. 
Then the gap equation and the pion pole condition assume
the forms \refb{gaa} and \refb{polen1} with 
${\overline \Pi}\rightarrow {\overline \Pi}_{LB}$.
In particular, the gap equation takes the explicit form
\be
1 - 24 {\overline G} \int_{LB} \frac{dx}{x} 
\int_{LB} \frac{d^2q_{\perp}}{\left(2\pi\right)^3} = \frac{{\overline m}}{M},
\label{gaaa}
\ee 
where the LB condition is given by \refb{cond}. Performing the
variable transformation \refb{trans}, we see that this is just the
gap equation in the 3M regularization scheme (eq. \refb{gapr})
with $G\rightarrow {\overline G}$ and $m\rightarrow {\overline m}$.

Let us then compare
a treatment based on the 3M regularization in terms of the usual
variables with the 'extended LB regularization' in terms of LC variables.
We treat the constituent
quark mass $M$ as a free parameter, choosing the same value in both
regularization schemes. Since the expressions for $f_{\pi}$ are the same
in the two treatments, we obtain the same value for $\Lambda$ in both
schemes by requiring that the experimental value for $f_{\pi}$ is
reproduced. Then the 4-fermi
coupling constant is determined by requiring that the experimental
pion mass is reproduced. In the 3M scheme this is expressed as
$1+2G \Pi_{3M}(m_{\pi}^2)=0$, while in the LB scheme we have 
$1+2{\overline G}\,\, {\overline \Pi}_{LB}(m_{\pi}^2)=0$. Since
${\overline \Pi}_{LB}=\Pi_{3M}$, this gives $G_{3M}={\overline G}_{LB}$.
Then the gap equation can be used in order to determine the value of
the current quark mass. As is clear from the above discussion following
eq. \refb{gaaa}, this gives ${\overline m}_{LB}=m_{3M}$. 
Therefore, the current quark masses ($m$) are different
in the two schemes, but also the quark condensates are different:
>From \refb{valid} and \refb{equal}, as well as from 
${\overline m}_{LB}=m_{3M}$ we have
\begin{eqnarray}
\langle {\overline \psi} \psi \rangle_{LB} &=&
\langle {\overline \psi} \psi \rangle_{3M}+M F_{LB} \label{relcond} \\
m_{LB}&=&\frac{m_{3M}}{1-2G_{3M}F_{LB}}.
\end{eqnarray}
As a consequence the GOR relation holds
in both schemes: In the 3M scheme we have eq. \refb{gor} with $m=m_{3M}$,
and following the same steps in the extended LB scheme we obtain
\be
-m_{LB} \langle {\overline \psi} \psi \rangle_{LB} = 
f_{\pi}^2 m_{\pi}^2
\left(1-\frac{m_{LB}}{M}\right) \left(1+m_{\pi}^2 C\right).
\label{gor2}
\ee
This 'invariance' of the product of the current quark mass and the
quark condensate holds also for  
$m \langle H| {\overline \psi} \psi |H \rangle$, where $|H\rangle$
denotes a hadronic state, and means that in actual calculations the
regularized form $F_{LB}$ and the value of $m$ in the LB scheme
are never needed: From the Feynman-Hellman theorem we have
${\displaystyle \langle H|{\overline \psi} \psi |H \rangle = 
\frac{\partial M_H}{\partial M} \frac{\partial M}{\partial m}}$. The 
second factor in the latter expression represents the 'vacuum contribution',
and its product with $m$ can be written in the form (see Appendix B)
\be
m \frac{\partial M}{\partial m} = \frac{m}{1-2G \Pi_{\sigma}(0)} =
\frac{{\overline m}}{1-2{\overline G}\,\, {\overline \Pi}_{\sigma}(0)},
\label{invar}
\ee
where $\Pi_{\sigma}(k)$ is the bubble graph in the $\sigma$ meson channel,
and $\Pi_{\sigma}(k)={\overline \Pi_{\sigma}}(k)-F$. 
>From our above discussion it is then clear that the product \refb{invar},
and therefore also the '$\sigma$-term' 
$m \langle H| {\overline \psi} \psi |H \rangle$, 
takes the
same value in the ET treatment based on the 3M cut-off and in the LC
theory based on the extended LB cut-off. 

The above discussion shows that, although the parameters $G$ and $m$
in the LC theory based on the extended LB cut-off scheme are different
from those of the ET theory based on the 3M cut-off scheme, the physical
predictions are the same. In this sense, the parameters ${\overline G}$
and ${\overline m}$ introduced in the LC theory by eqs. \refb{geff} and
\refb{meff} can be considered as 'renormalized' parameters, which take the
same values as $G$ and $m$ in the ET theory. (The actual value of the
'renormalization constant' $1+2GF$ is never needed.) 

We conclude that the extended LB scheme is physically equivalent 
to the 3M scheme.  

\subsubsection{Transverse (TR) cut-off scheme}

In this scheme \cite{TRAN}, one notes that the
divergence of the bubble graph ${\overline \Pi}$ with respect to the
longitudinal momentum integration is logarithmic (see \refb{pibarx} and
\refb{pibar}), and
therefore one subtraction is sufficient to make it finite. As a
consequence, the following equations are already finite
with respect to the longitudinal momentum integration: (i) The pion 
pole condition written in the form \refb{polen1}, (ii) the pionic t-matrix
\refb{taub} written in the form
\be
{\overline \tau}(k)=\frac{-i}{{\overline \Pi}(k)-{\overline \Pi}(m_{\pi}^2)},
\label{reno}
\ee 
(iii) the pion decay constant, and (iv) the quark distribution function.
The TR scheme now consists in introducing a transverse momentum cut-off
$\Lambda_{TR}$ in these equations, while the remaining divergent
equations (the gap equation \refb{gaa}, and eq. \refb{exf} determining 
the quantity $F$) are not used explicitly. 

Note that in this method the gap equation is used only to formally 
rewrite the pion 
pole condition into the form \refb{polen1}, which is finite with respect to the
longitudinal momentum integration. It is easy to show that this
method works also for the sigma meson channel, i.e; the gap equation can be
used to rewrite the equation for the sigma meson mass into a form which
is finite with respect to the longitudinal momentum integration. 
It is a disadvantage of this method, however, that it works strictly
only within the present $1/N$ expansion scheme: The gap equation and
the eigenvalue equation for the meson mass must be of the same order in
$1/N$. For the lagrangian \refb{lag}, this is true only for the
pion and sigma meson, since the interaction in other mesonic channels
is due to the exchange term and suppressed by one power of $1/N$.
It is, nevertheless, possible to apply the TR scheme also to these cases
by performing a kind of mass renormalization, that is to use the eigenvalue 
equation for the meson mass only in a formal sense to rewrite the
t-matrix into a renormalized
form analogous to \refb{reno}, which is finite with respect to the
longitudinal momentum integration. Such a renormalization procedure,
in which the gap equation and the eigenvalue equations for the hadron
masses are used only in a formal sense, is quite different in spirit
from the common usage of the NJL model.

\section{Numerical results for the quark LC momentum distribution in the pion}
\setcounter{equation}{0}

In this section we present numerical results for the  quark LC momentum 
distribution
\refb{result} 
regularized by using the two regularization schemes discussed in
subsect. 3.5, i.e; the extended LB scheme and the TR 
scheme. We choose the constituent quark mass $M=300 MeV$ in both cases. Then
by requiring that \refb{fpir} (or \refb{fpil} with the condition \refb{cond})
reproduces the experimental
value $f_{\pi}=93 MeV$, we obtain for the LB cut-off $\Lambda=1.47 GeV$
(corresponding to a 3-momentum cut-off of $\Lambda_3=0.67 GeV$). 
\footnote{For the LB scheme, we obtain $G_{3M}={\overline G}_{LB}=
4.56 GeV^{-2}$ from the pion pole condition (using $m_{\pi}=140 MeV$), 
and $m_{3M}={\overline m}_{LB}=5.3 MeV$ from the gap equation.}   
Similarly,
for the TR scheme we use \refb{fpil} with $|{\bold q}_{\perp}|<\Lambda_{TR}$
to obtain $\Lambda_{TR}=0.48 GeV$. 
The value of the pion-quark coupling constant, defined at 
$k^2=m_{\pi}^2=(140 MeV)^2$, becomes $g=3.15$ ($3.16$) in the LB (TR) scheme.
(These values are close to each other, since in the chiral limit we
have $g=M/f_{\pi}=3.23$ in both schemes.) 

The resulting distributions multiplied by $x$ are shown by 
the dashed lines
in fig. \ref{lcd1} for the LB scheme, and in fig. \ref{lcd2}
for the TR scheme. 
Due to the condition \refb{cond} in the LB regularization, the 
distribution $q_v(x)$ is nonzero only in the interval
$x_1<x<x_2$ and tends to zero smoothly at the end points $x_{1,2}$.
In the TR regularization, on the contrary, $q_v(x)$  extends over the
whole interval $0\leq x \leq 1$ and is nonzero at the end points.
In this case, due to the smallness of the pion mass, $q_v(x)$ shows
only little variation as a function of $x$, which leads to the almost
linear behaviour of $x q_v(x)$ shown in fig. \ref{lcd2}. This is very 
similar to the result of a previous calculation \cite{PIONS} 
using the Pauli-Villars
regularization based on Feynman integral representation of $q_v(x)$
(see the discussion below eq.\refb{ide}).    

To connect these results with experiment, we assume as usual \cite{JAFFE}
that our
NJL results are valid at some low-energy scale $Q_0^2$, and perform the
$Q^2$ evolution using the DGLAP equations \cite{AP} up to the
value $Q^2=4 GeV^2$, where parametrized fits to the experimental
results are available \cite{EMP}. We used the computer code of ref.
\cite{SAGA} to solve the DGLAP equations up to the next-to-leading order,
using the parameters $\Lambda_{QCD}=0.25 GeV$ and $N_f=3$. The value of
$Q_0^2$ is considered here as a parameter to be adjusted such as to 
reasonably reproduce the experimental distribution. In both regularization
schemes the value obtained in this way is $Q_0^2=0.18 GeV^2$, i.e;
$Q_0$ is of the order of the constituent quark mass. 

The results of the $Q^2$ evolution are shown by the solid lines
in figs. \ref{lcd1} and \ref{lcd2}, 
\footnote{Since the computer code used for the $Q^2$ evolution \cite{SAGA}
requires an input distribution $xq_v(x)$ which vanishes at $x=1$, we
artificially modified it for $x$ very close to $1$ such that it goes like
$(1-x)^n$ with some power $n$. ($n=10$ was used in the actual calculation.)}
and are compared to the empirical 
distribution
given in ref. \cite{EMP} (dotted lines). The fact that in the LB scheme the 
theoretical
distribution is too low for high $x$ is connected to the LB condition 
\refb{cond}
which gives a vanishing input distribution for high $x$ (see the
dashed line in fig. 4). The use of smaller values for $Q_0^2$ would lead to
better agreement around the peak position, but to a larger discrepancy
for higher $x$. A similar situation is encountered also in the case
of the nucleon structure function \cite{MINEO}. On the other hand, the result 
based on the TR scheme reproduces the empirical distribution almost exactly.  
Therefore, from the phenomenological point of view, a regularization
scheme which does not restrict the longitudinal momenta seems to be
favorable.  

\section{Summary and conclusions}

In this work we investigated the NJL model in the LC quantization and
its relation to the usual ET formalism, putting special emphasis on the 
application to the pion structure function. We first reviewed the NJL model
and the calculation of the pion structure function in the ET formalism.
We showed that the Lepage-Brodsky (LB) cut-off scheme for
the quark LC momentum distribution can be applied consistently in
connection with the covariant 3-momentum cut-off (3M) scheme
for the gap equation and the $q {\overline q}$ polarization propagator.

The introduction of LC variables in the ET formalism is simply a
transformation of variables. As we discussed in the introduction, however,
the interpretation of the structure function as LC momentum distributions
of the constituents in the ground state target
is most natural in the LC quantization procedure.  
Due to this fact, and other physically appealing features of the LC
formalism, we formulated the LC quantization
of the NJL model. The essential part was the solution of the fermionic
constraint equation using the $1/N$ expansion up to the next-to-leading
order. We showed that, up to this order, the interaction
lagrangian on the LC for the dynamical component of the fermion field
has the same form as obtained by using the free field equations to
solve the fermionic constraint, but the original 4-fermi coupling constant
is replaced by an effective (renormalized) one. The Feynman rules 
for the LC lagrangian are very similar to
those for the ET lagrangian: The propagators for internal lines 
differ just by a contact term, and 
the 4-fermi vertices differ by a multiplying factor. For physical
quantities these differences formally cancel with each other, leading to
the same results as in the ET theory.
We have demonstrated this equivalence for the eigenvalue equation 
determining the pion mass, the pion decay 
constant, and the pion structure function in the Bjorken limit. 
In the LC quantization, these quantities can be
evaluated either
by using the language of Feynman diagrams as described above, or by using 
the Fock space 
expansions based on the hamiltonian formalism. Both methods lead to
the same results.

In this work, we avoided the difficulty of regularizing tadpole-like
diagrams in the LC theory by relating them to 2-point Green functions 
at zero momentum. 
We then formulated two possible regularization schemes: The first one
was an extension of the "invariant mass cut-off scheme" \cite{LB}.
We have shown that this scheme leads to the same physical predictions as
the covariant 3-momentum
cut-off (dispersion cut-off) scheme in the ET formulation, provided that
in both treatments the parameters are fixed such as to have the same values
for $M,\,f_{\pi},$ and $m_{\pi}$. The application of this scheme 
to problems others than studied in this paper, like the nucleon structure
function, is straight forward \cite{MINEO}. The second one 
("transverse cut-off scheme")
employs a cut-off only for the transverse momentum components. In this
scheme, the gap equation is used only to formally rewrite the eigenvalue
equation for the pion and sigma mass into a form which is finite with 
respect to the longitudinal momentum integrations. We noted that
this method works only
within the present $1/N$ expansion scheme, and to apply it to other
meson channels or to the nucleon, one has
in addition to perform a kind of mass renormalization. Such a
procedure, however, departs significantly from the common 
usage of the NJL model.

As an application, we evaluated the quark LC momentum distribution in the
pion numerically and performed the $Q^2$ evolution to compare the
results with parametrizations based on experimental data.
Since in the LB regularization scheme the input distribution is zero
in an interval near the end points, the slope of the
evolved distribution on the large $x$ side is somewhat too large compared 
with the empirical distribution. Nevertheless, a fair agreement
can be achieved by assigning a low energy scale $Q_0^2=0.18 GeV^2$
to the NJL model, which is similar to the values used in previous
investigations of the pion structure function. On the other hand, 
in the TR regularization scheme one can reproduce the empirical
distribution almost exactly using the same value of $Q_0^2$. This indicates
that for phenomenological applications a regularization scheme which
does not restrict the longitudinal momentum components seems to be
preferable.

\vspace{1cm}
{\sc Acknowledgements}

The authors would like to thank M. Miyama and S. Kumano for the
computer program used for the $Q^2$ evolution (ref. \cite{SAGA}).
One of the authors (W.B.) is grateful to A. W. Thomas and A. W. Schreiber
for discussions on the pion structure function, and to M. Burkardt and 
K. Itakura for discussions on light cone theories.


\newpage

\appendix
{\LARGE\bf Appendices}

\section{LC variables and Dirac matrices}
\setcounter{equation}{0}
In this Appendix we summarize our conventions for LC variables and Dirac
matrices, and give some useful identities which have been used in the
main text.
We use the LC variables $a^{\pm}=\frac{1}{\sqrt{2}}\left(x^0 \pm x^3\right)$,
$a_{\pm}=\frac{1}{\sqrt{2}}\left(x_0 \pm x_3\right)$, and
$x_{\perp}^i=-x_{\perp\,i}$ ($i=1,2$). Then we have for the Lorentz scalar
product 
$p\cdot x=p_+x^+ + p_-x^- - {\bold p}_{\perp}\cdot {\bold x}_{\perp}$.
In the LC quantization procedure, $x^+$ is the 'time' variable and 
$x^-,{\bold x}_{\perp}$ are the 'space' variables. Similarly,
$p_+$ is the 'energy' variable and 
$p_-,{\bold p}_{\perp}$ are the 'momentum' variables. The dispersion
relation for a free particle with mass $M$ is 
$p_+ \equiv e_p = \left({\bold p}_{\perp}^2+M^2\right)/2p_-$, 
see eq. \refb{eps}.
For the derivatives we use the conventions
$\partial_{a}=\partial/ \partial x^{a}$ ($a=+,-,1,2$).

Analogous definitions hold for the LC Dirac matrices 
$\gamma^{\pm}, {\bold \gamma}_{\perp}$ in terms of the $\gamma^{\mu}$.
It follows that $\fslash{\partial}=\gamma^+ \partial_+ +
\gamma^- \partial_- - {\bold \gamma}_{\perp}\cdot {\bold \partial}_{\perp}$.   
Note that $\gamma^+ \gamma^+=\gamma^- \gamma^-=0$. The projection
operators $\Lambda_{\pm}$ have been introduced in eq. \refb{proj}.
Equivalent useful forms are $\Lambda_{\pm}=\frac{1}{\sqrt{2}}
\gamma^0 \gamma^{\pm}= \frac{1}{\sqrt{2}}
\gamma^{\mp} \gamma^0 = \frac{1}{2} \left(1 \pm \gamma^0 \gamma^3\right)$.
We have the following useful identities:
\begin{eqnarray}
\gamma^- \Lambda_+ = \Lambda_+ \gamma^+ 
= \gamma^+ \Lambda_- = \Lambda_- \gamma^-=0 \label{one} \\  
\Lambda_{\pm} \gamma^0 = \gamma^0 \Lambda_{\mp} \label{two}.  
\end{eqnarray}   

For the quark bilinears it follows that
\begin{eqnarray}
{\overline \psi} \psi &=& \psi_+^{\dagger} \gamma^0 \psi_- + 
\psi_-^{\dagger} \gamma^0 \psi_+ \\
{\overline \psi} \gamma^{\pm} \psi &=& \sqrt{2} \psi_{\pm}^{\dagger}
\psi_{\pm} \\
{\overline \psi} {\bold \gamma}_{\perp}\cdot {\bold \partial}_{\perp}
\psi &=&  \psi_+^{\dagger} {\bold \alpha}_{\perp}\cdot {\bold \partial}_{\perp}
\psi_- + \psi_-^{\dagger} {\bold \alpha}_{\perp}\cdot {\bold \partial}_{\perp}
\psi_+\,\,. 
\end{eqnarray}

We also give some identities which were used in deriving eqs. \refb{ide1}
and \refb{final}:
\begin{eqnarray}
\Lambda_{\pm} \left(\fslash{p} + M\right) \Lambda_{\pm} &=&
\Lambda_{\pm} \left(-{\bold \gamma}_{\perp}\cdot {\bold p}_{\perp} + M \right)
\\
\Lambda_{\pm} \left(\fslash{p} + M\right)\gamma^0 \Lambda_{\pm} &=& 
\sqrt{2} \Lambda_{\pm} p_{\mp} \\
\Lambda_{\pm} \left(\fslash{p} + M\right) \Lambda_{\mp} &=&
\gamma^{\mp} p_{\mp} \\
\Lambda_{\pm} \left(\fslash{p} + M\right)\gamma^0 \Lambda_{\mp} &=&
\frac{1}{\sqrt{2}} \gamma^{\mp} 
\left({\bold \gamma}_{\perp}\cdot {\bold p}_{\perp} + M \right)\,\,.
\end{eqnarray}

The Dirac spinors are normalized according to 
$u^{\dagger}({\bold p},s') u({\bold p},s)=2 E_p \delta_{s' s}$ and
$v^{\dagger}({\bold p},s') v({\bold p},s)=2 E_p \delta_{s' s}$.
The spinors $u_{+} = \Lambda_{+} u$ and
$v_{+} = \Lambda_{+} v$ then satisfy the orthogonality and completeness
relations
\begin{eqnarray}
u_+^{\dagger}({\bold p},s') u_+({\bold p},s) &=&  
v_+^{\dagger}({\bold p},s') v_+({\bold p},s) = \sqrt{2} p_-\,\delta_{ss'} \\
\sum_s u_+({\bold p},s) u_+^{\dagger}({\bold p},s) &=&
\sum_s v_+({\bold p},s) v_+^{\dagger}({\bold p},s) = \sqrt{2} \Lambda_+ p_-
\,\,.
\end{eqnarray}

\section{Relations between the condensate and 2-point functions}
\setcounter{equation}{0}

We first derive relation \refb{gene} in the ET formalism 
from the chiral
Ward identity for the pion decay vertex. If we define
$J_A^{i\,\mu}={\overline \psi} \gamma^{\mu} \gamma_5 \tau_i \psi$
and $\pi^i = {\overline \psi}\, i\gamma_5 \tau_i \,\psi$, the chiral
Ward identity follows as usual from the PCAC relation
$\partial_{\mu} J_A^{i\, \mu} = 2m\, \pi^i$:
\begin{eqnarray}
\lefteqn{\partial_{\mu} \langle 0 | T \left( J_A^{i\,\mu}(x) \pi^j(x') 
\right) |0\rangle} \nonumber \\
&=& 2m \langle 0 | T \left( \pi^i(x) \pi^j(x') \right) |0\rangle 
-2i \delta_{ij} \delta^{(4)}(x'-x) \langle 0| {\overline \psi}\psi
|0 \rangle.
\label{ward}
\end{eqnarray}
To lowest order in $1/N$, the correlation function 
$\langle 0 | T \left( \pi^i(x) \pi^j(x') \right) |0\rangle$ is given by the
series of bubble graphs. Considering the Fourier transform of \refb{ward}
for $k=0$, we therefore obtain in the leading order of $1/N$
\be
0 = 2im \frac{\Pi(0)}{1+2G\Pi(0)} -2i\langle 0| {\overline \psi}\psi
|0 \rangle.  
\label{kzero}
\ee
Since the constituent quark mass is defined by 
$M=m-2G \langle 0| {\overline \psi}\psi |0 \rangle$, this relation
can be written in the form $\langle 0| {\overline \psi}\psi
|0 \rangle = M \Pi(0)$, which is eq. \refb{gene}.

Next we derive eq. \refb{invar} from the general relation
\be
\frac{\partial \langle {\overline \psi}\psi \rangle}{\partial M} =
- \Pi_{\sigma}(0),
\label{relat}
\ee
where $\Pi_{\sigma}(k)$ is the bubble graph in the $\sigma$ meson
channel.  Eq. \refb{relat} can be derived most easily by using the
functional integral. It is easy to check that this relation holds
in the 3M regularization scheme to leading order in $1/N$, i.e;
${\displaystyle \frac{\partial \langle {\overline \psi}\psi 
\rangle _{3M}}{\partial M} =
- \Pi_{\sigma,3M}(0)}$. Then, in our extended LB scheme, differentiating
\refb{relcond} with respect to $M$ and requiring that the result
is equal to $-\Pi_{\sigma,LB}(0)=-({\overline \Pi}_{\sigma,LB}(0)-F_{LB})$
leads to the requirement that $F_{LB}$ is independent of $M$. 
By differentiating
the gap equation \refb{gap} with respect to $M$ we obtain
\be
m \frac{\partial M}{\partial m} = \frac{m}{1+2G \frac{\partial}{\partial M}
\langle {\overline \psi}\psi \rangle } =   
\frac{m}{1-2G \Pi_{\sigma}(0)} = \frac{\overline m}{1-2{\overline G}\,\, 
{\overline \Pi}_{\sigma}(0)},
\ee
where we used ${\displaystyle \frac{G}{\overline G}=\frac{m}{\overline m}=
\frac{1}{1-2{\overline G} F}}$. This proves eq. \refb{invar}.  

\section{Bound state equation}
\setcounter{equation}{0}

In this Appendix we derive the bound state equation in the TDA 
for the LC wave function
in the pionic $q {\overline q}$ channel, using (i) 
the TDA equation in the hamiltonian formalism,
and (ii) the BS equation.

\subsection{Hamiltonian formalism}

The LC hamiltonian derived from the lagrangian eq. 
\refb{end} is given in terms of $\psi_+$ by
\begin{eqnarray}
H &=& \int dx^- d^2x_{\perp} \left( \psi_+^{\dagger} \left(-
{\bold \partial}_{\perp}^2+M^2\right) \frac{1}{\sqrt{2}i\partial_-} \psi_+
\right. \nonumber \\
&+& \left. {\overline G} \left[:\psi_+^{\dagger} \gamma_5 \gamma^0 {\bold \tau}
\left\{ \frac{-1}{\sqrt{2} i \partial_-} \left(i{\bold \alpha}_{\perp}
\cdot {\bold \partial}_{\perp} + M\right) + \left(i{\bold \alpha}_{\perp}
\cdot \stackrel{\leftarrow}{\bold \partial}_{\perp} 
- M\right)\frac{1}{\sqrt{2} i \stackrel{\leftarrow}{\partial}_-}
\right\} \psi_+ :\right]^2 \right) \nonumber \\
&+& \sigma\,\, {\rm channel},
\label{ham}
\end{eqnarray}
where we show only the interaction in the pionic channel explicitly. 
We now insert \refb{mod} into the above hamiltonian and
apply it to the pion state \refb{fock}, keeping only those terms 
which are again of the form \refb{fock}. (This corresponds to the TDA, and
means that in the interaction hamiltonian only terms of the form 
$b^{\dagger}\,d^{\dagger}\, b\, d$ are kept.)
The result is expressed as an eigenvalue equation for the 'energy' 
$\epsilon_k=({\bold k}_{\perp}^2+m_{\pi}^2)/2k_-$, i.e;
$H |k\rangle = \epsilon_k | k\rangle $. The procedure is
tedious but straight forward, and therefore we will give only the final
result. One ends up with a TDA equation, where the kernel has both
direct and exchange terms. The kernel for the direct term leads to a color
factor $3$, and in order to be consistent with the leading order in the
$1/N$ expansion used throughout this paper we keep only the direct term.
The result is
\begin{eqnarray}
\lefteqn{\left(\epsilon_k - \frac{{\bold q}_{\perp}^2+M^2}{2k_-x}
- \frac{\left({\bold k}-{\bold q}\right)_{\perp}^2+M^2}{2k_-(1-x)} \right)
\Phi^{(S)i}_{k,a_1 a_2}({\bold q})} \nonumber \\
&=&-\frac{\overline G}{6 k_-^3} \int_0^1 \frac{dx'}{\sqrt{x(1-x)x'(1-x')}}
\int \frac{d^2q_{\perp}'}{\left(2\pi\right)^3} 
\left(u_+^{\dagger}({\bold q}, s_1)\chi^{\dagger}(t_1) {\hat X}_k^i({\bold q}) 
v_+({\bold k}-{\bold q},s_2) \right. \nonumber \\
&\times& \left. \chi(t_2) \delta_{c_1 c_2} \right) 
\left(v_+^{\dagger}({\bold k}-{\bold q}', s_2')\chi^{\dagger}(t_2') 
{\hat X}_k^j({\bold q}') u_+({\bold q}',s_1') \chi(t_1') \delta_{c_1' c_2'}
\right)\Phi^{(S)}_{k,a_1' a_2'}({\bold q}') \nonumber \\
\label{hami}
\end{eqnarray}
with (see eq. \refb{vert})
\be
{\hat X}_k^i({\bold q})= k_- X_k^i({\bold q})= \gamma_5   \tau_i \left(
\frac{{\bold \gamma}_{\perp}\cdot{\bold q}_{\perp}+M}{\sqrt{2}x}
+\frac{{\bold \gamma}_{\perp}\cdot\left({\bold q}-{\bold k}\right)_{\perp}+
M}{\sqrt{2}(1-x)}\right).
\label{defaa}
\ee
Here the variables $x$ and $x'$ have been introduced according to
$q_-=k_- x$ and $q_-'=k_- x'$. The notation is as in eq. \refb{final},
i.e; $a_i=(s_i,t_i,c_i)$.
It is clear that for ${\bold k}_{\perp}=0$ the solution of this 
equation is of the form \refb{final}. By inserting this form back into
\refb{hami} we get the eigenvalue equation for $\epsilon_k$ which
coincides with the equation $1+2{\overline G}\,\, {\overline \Pi}(k)=0$
with ${\overline \Pi}(k)$ given by eq. \refb{pibar}.

\subsection{BS equation}
The BS wave function for the dynamical components of the fermion field
has been defined in eq.\refb{dyn} and satisfies the following BS equation:
\be
{\tilde \Phi}^i_{k,\alpha \beta}(q) = \frac{2 i {\overline G}}{k_-^2}
\int \frac{d^4q'}{\left(2\pi\right)^4}
\left({\overline S}_{++}(q) {\hat X}_k^i({\bold q}) 
{\overline S}_{++}(q-k)\right)
_{\alpha \beta} \,\frac{1}{3} \,\,Tr \left({\hat X}_k^j({\bold q}')
{\tilde \Phi}^j_{k}(q')\right),
\label{bs}
\ee
where the propagator ${\overline S}_{++}$ is given in eq. \refb{pp},
and the interaction kernel \refb{vert} has been expressed in terms
of the quantity ${\hat X}$ of \refb{defaa}, 
where $q_-=k_- x$ and $q_-'=k_- x'$.
We integrate eq. \refb{bs} over $q_+$ to derive the equation for the
wave function $\Phi_k^{(S)}({\bold q})$ defined in eq. \refb{sch}:
\begin{eqnarray}
\lefteqn{\left(k_+ - \frac{{\bold q}_{\perp}^2+M^2}{2k_-x}
- \frac{\left({\bold k}-{\bold q}\right)_{\perp}^2+M^2}{2k_-(1-x)}\right)
\Phi^{(S)i}_{k,\alpha \beta}({\bold q})} \nonumber \\
& =& - \frac{{\overline G}}{k_-}
\int_0^1 dx' \int \frac{d^2 q'_{\perp}}{\left(2\pi\right)^3}
{\hat X}^i_{k,\alpha \beta}({\bold q}) \, \frac{1}{3} Tr \left( 
{\hat X}_k^j({\bold q}')
\Phi^{(S)j}_{k}({\bold q}')\right).
\end{eqnarray}  
The corresponding wave function in the spin representation (see the
discussion below eq. \refb{sch})
\be
\Phi^{(S)i}_{k,a_1 a_2}({\bold q})= \frac{1}{k_- \sqrt{2x(1-x)}}
u_+^{\dagger}({\bold q},s_1) \chi^{\dagger}(t_1) 
\Phi^{(S)i}_{k}({\bold q}) v_+({\bold k}-{\bold q},
s_2) \chi(t_2) \delta_{c_1 c_2}
\ee
then satisfies the equation \refb{hami}.

\section{Bound state matrix elements using the LC BS wave functions}
\setcounter{equation}{0}

If we use the BS wave function ${\overline \Phi}_k$ in the diagram 
for pion decay, referring
to the $+$ component of the axial vector current, the
resulting expression differs from eq. \refb{loop} only by the replacement
$S\rightarrow {\overline S}$ for the two quark lines.  
(Note that the $+$ component of the axial vector current can be expressed
as ${\overline \Psi} \gamma^+ \gamma_5 {\bold \tau} \Psi$.)
Due to relation \refb{sbar} it is then easy to see that this replacement
leads to no changes. If we use the wave function ${\tilde \Phi}_k$, the trace
factor involves two extra factors $\Lambda_+$, but due to the relation
$\gamma^0 \gamma^+ = \sqrt{2} \Lambda_+$ these factors can be left out,
leading again to the same result.  

If we use the BS wave function ${\overline \Phi}_k$ together with the
propagators ${\overline S}$ in the diagram fig. \ref{dia} for the
pion structure function, the
resulting expression differs from eq.\refb{loop1} only by the replacement
$S\rightarrow {\overline S}$ for all three quark lines. From \refb{sbar}
and the familiar relation $S(q)\gamma^0 \gamma^+ S(q) =
- \partial S(q)/\partial q_+$ it follows that also
${\overline S}(q)\gamma^0 \gamma^+ {\overline S}(q) =
- \partial {\overline S}(q)/\partial q_+$, and performing a partial
integration in $q_+$ we end up with       
$q_v(x)=-g^2 \partial {\overline \Pi}(k,x)/ \partial k^2$.
Due to relation \refb{piece} this is the same as \refb{qv}.
If we calculate the diagram fig. \ref{dia} by using the wave function
${\tilde \Phi}_k$ and the propagators \refb{pp}, the trace factor is
\footnote{The quantity ${\overline S}_{++}^{-1}$ in eqs. \refb{trace},
\refb{last} is understood as the inverse of eq. \refb{pp} except for
the projection operator $\Lambda_+$.}
\be
Tr\left[\Lambda_+ {\overline S}(q)\gamma^0 \gamma_5 {\overline S}(q-k)
\Lambda_+ {\overline S}_{++}^{-1}(q-k) \Lambda_+ {\overline S}(q-k) \gamma^0
\gamma_5 {\overline S}(q) \Lambda_+ \gamma^0 \gamma_5\right].
\label{trace}
\ee
Using the identity (see eq. \refb{defz} and \refb{ide1})
\be
{\overline S}(q-k)\left(
\Lambda_+ {\overline S}_{++}^{-1}(q-k) \Lambda_+ {\overline S}(q-k)\right)
={\overline S}(q-k)\Lambda_+\left(1+Z({\bold q}-{\bold k})\right)
={\overline S}(q-k)
\label{last}
\ee
and $\gamma^0 \gamma^+ = \sqrt{2} \Lambda_+$, the result is the same as
obtained by calculating the diagram fig. \ref{dia} by using the propagators
${\overline S}$. As was shown above, this gives the result \refb{qv}
of the ET theory.

\newpage

\section*{Figure captions}

\begin{enumerate}

\item {External momenta for the $q {\overline q}$ t-matrix.}
\label{mom}

\item {Diagram for the quark LC momentum distribution in the pion.
The double line represents the pion, the shaded area the pionic
vertex function, and cross represents the external operator
$\gamma^+$.}
\label{dia}

\item{Vertex correction to the 'bare' vertex (the cross in fig. 
\ref{dia}).} 
\label{vertex}

\item {LC momentum distributions of the valence quark in the pion, using
the LB regularization scheme.
The dashed line shows the NJL model result (eq.\refb{result} with the
condition \refb{cond}) with the parameters
$M=0.3 GeV$ and $\Lambda=1.47 GeV$. The solid line shows the result
obtained by the QCD evolution up to next-to-leading order from 
$Q_0^2=0.18 GeV^2$ to $Q^2=4 GeV^2$, using $\Lambda_{QCD}=0.25 GeV$
and $N_f=3$. The dotted line shows the parametrization at $Q^2=4 GeV^2$ 
obtained in ref. \cite{EMP} by analyzing the experimental data.}
\label{lcd1}

\item {LC momentum distributions of the valence quark in the pion, using
the TR regularization scheme. The dashed line is calculated by using
the cut-off $\Lambda_{TR}= 0.48 GeV$ for the transverse
momentum integration in eq. \refb{result}. 
For further explanation, see the caption to fig. \ref{lcd1}.}
\label{lcd2}

\end{enumerate}

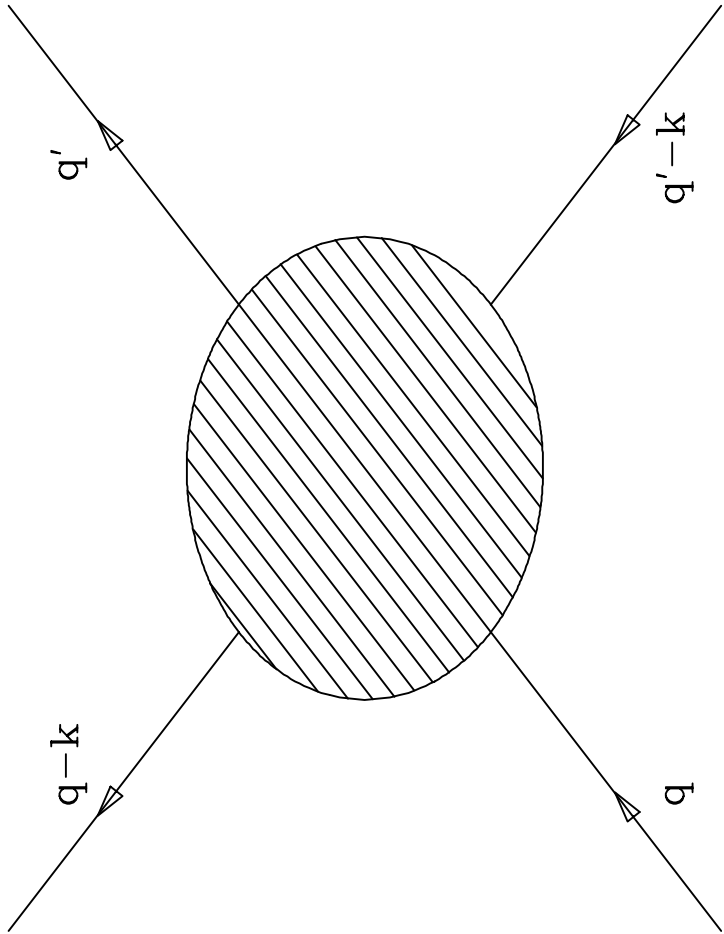
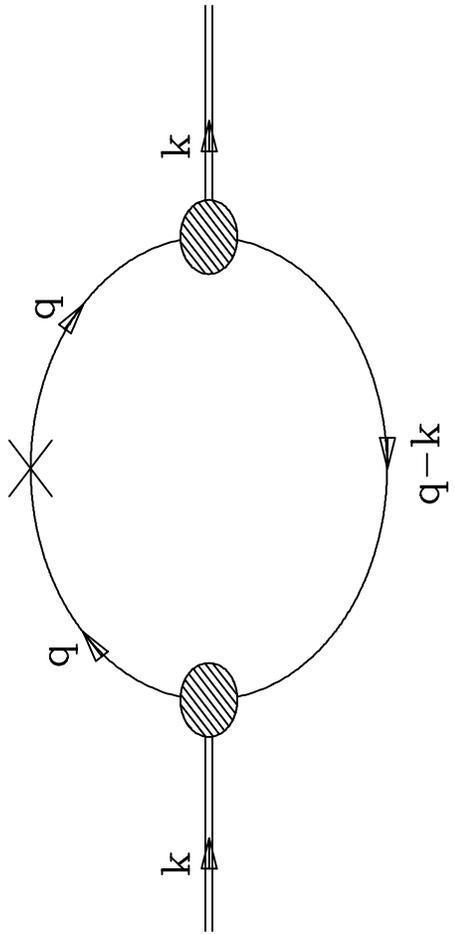
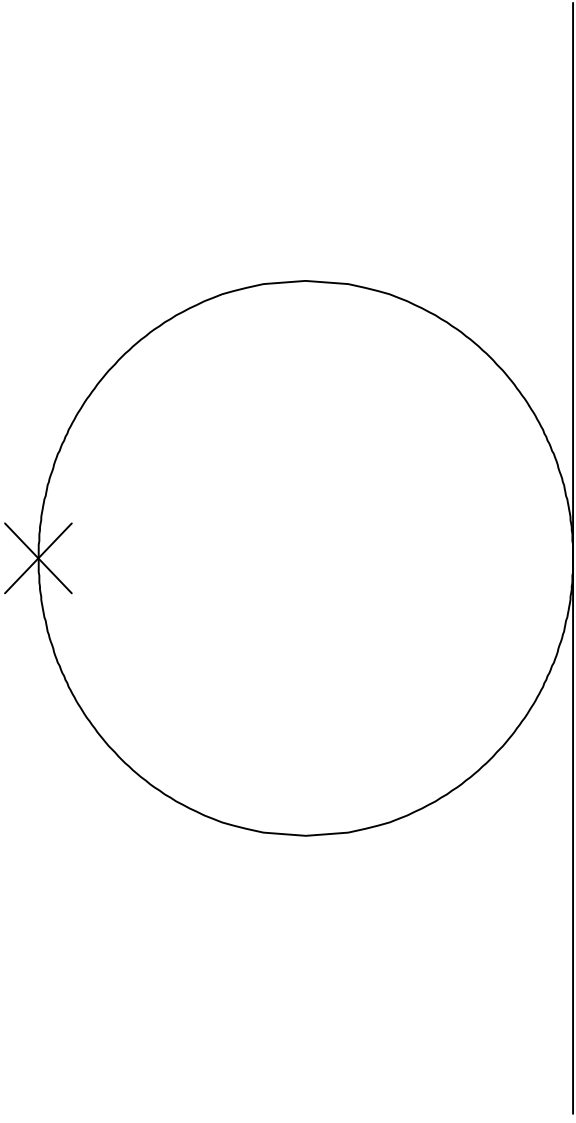
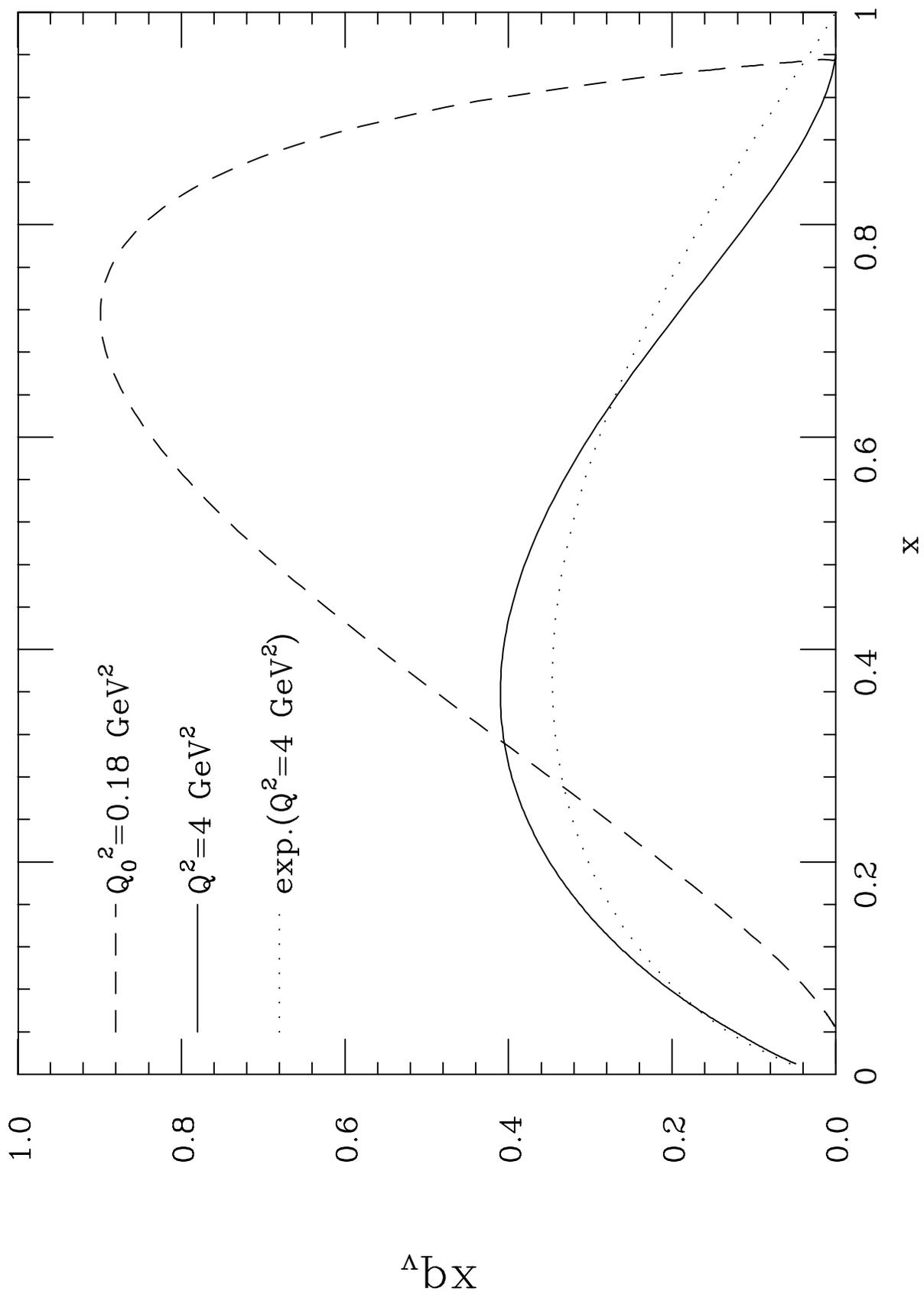
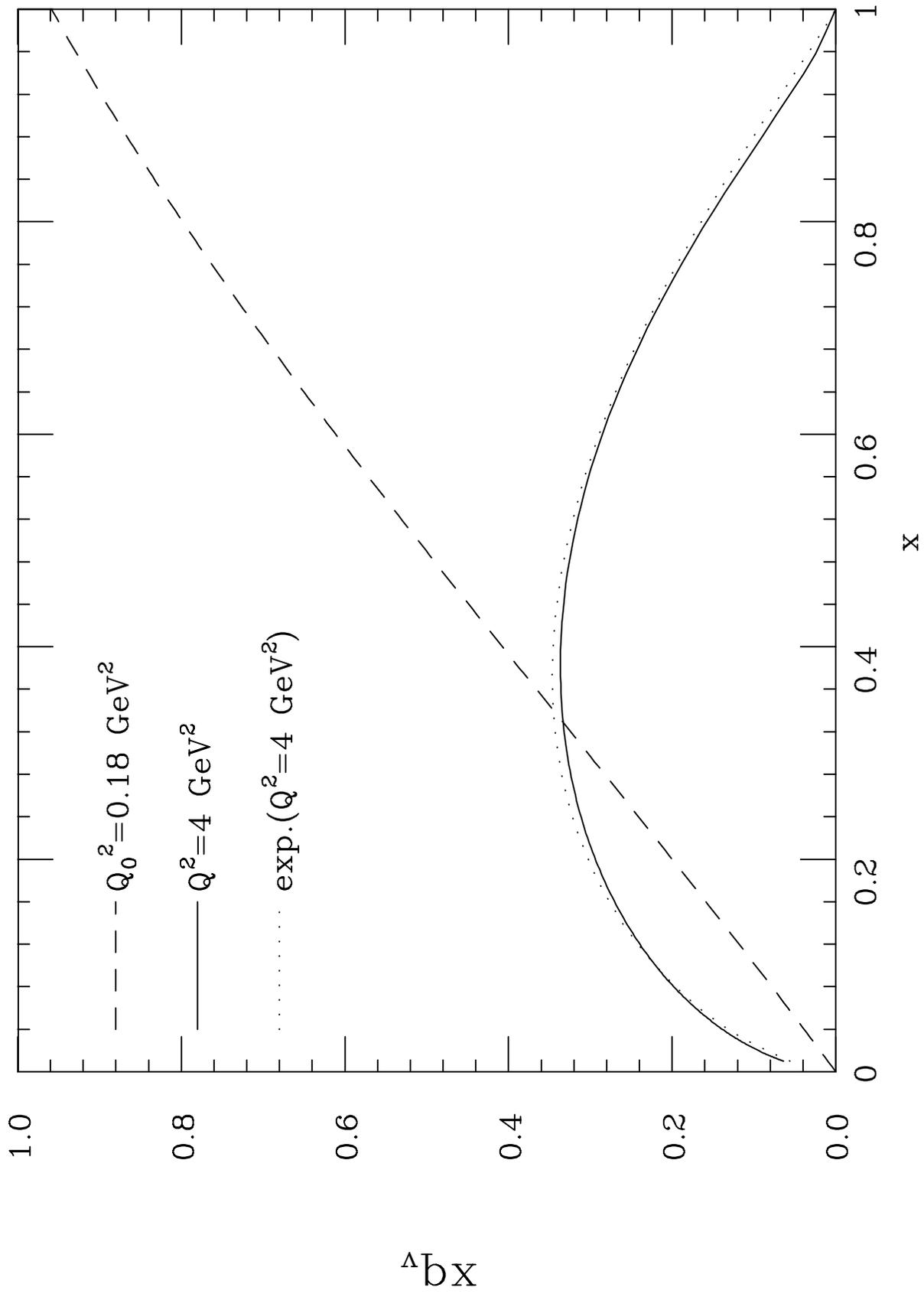
\end{document}